%% file: paper.tex
\documentclass[12pt,letterpaper]{article}
\usepackage[left=2cm,right=2cm,top=20mm,bottom=2cm]{geometry}\textheight=235mm
\usepackage[utf8]{inputenc}
\usepackage[english]{babel}
\usepackage{amsmath,amsfonts,amssymb,amsthm}
\usepackage[affil-it]{authblk}
\usepackage{graphicx,paralist}
\usepackage{bbm}
\usepackage{color}
\usepackage{subfigure}
\usepackage[format=hang,justification=raggedright]{caption}
\usepackage{datetime}
\usepackage{physics}
\newcommand{\ci}{\ensuremath{\mathbbm{i}}}
\newcommand{\psilabel}{\psi_{\mathfrak{L}}}

\input{hudef-b}

\def\hualst#1{\begin{align*} #1 \end{align*}}

\newcommand{\bci}{\begin{compactitem}}\newcommand{\eci}{\end{compactitem}}
\newcommand{\bcen}{\begin{compactenum}}\newcommand{\ecen}{\end{compactenum}}
\def\ds{\displaystyle}

\begin{document}

%\preprint{APS/123-QED}
%
\title{Life time of topological coherent modes of a Bose--Einstein condensate in a gravito optical surface trap}

\author[1]{Želimir Marojević}
\author[1]{Ertan Göklü}
\author[2]{Hannes Uecker}
\author[1,3]{Claus Lämmerzahl}
\affil[1]{
 ZARM - Center of Applied Space Technology and Microgravity\\
 D-28359 Bremen, Germany
}
\affil[2]{
 Institut für Mathematik, Carl von Ossietzky Universität Oldenburg\\
 D-26111 Oldenburg, Germany
}
\affil[3]{
 Institut für Physik, Carl von Ossietzky Universität Oldenburg\\
 D-26111 Oldenburg, Germany
}

\maketitle

\begin{abstract}
We give numerical estimates of various unstable stationary solutions of the Gross--Pitaevskii equation in an axially symmetric set up with a linear trapping potential along the symmetry axis, and a quadratic trapping along the radial direction. These represent topological coherent modes of Bose--Einstein condensates in a gravito optical surface trap (GOST). 
Despite their instability, we find that many of these solutions decay sufficiently slow, so that they could be realized experimentally.

\end{abstract}

\section{Introduction}
\def\Om{\Omega}

One way to study the behaviour of quantum matter in a gravitational field is the use of interferometry with neutrons, thermal atoms, or Bose-Einstein condensates. Another possibility is the study of eigenstates of matter, which has already been conducted with ultra cold neutrons falling down from different initial heights. The experiment in \cite{wallis_trapping_1992,abele_quantum_2003} has confirmed that the probability to find a neutron at a specific height is non classical and corresponds to the eigenfunctions of the Hamilton operator with a linear potential, which are the Airy functions. However, experiments with ultra cold neutrons are challenging due to a high loss rate. 

Here we propose to use Bose-Einstein condensates trapped in a gravito optical trap (GOST), composed of a reflecting surface realized by evanescent mirrors \cite{perrin_diffuse_2006}, and a dipole trap for the radial confinement.  

The most challenging part is to prepare initial states, which are not necessarily ground states. We think of states which possess more structure. These coherent topological modes could be realized with quantum control techniques \cite{bucker_twin-atom_2011,bucker_vibrational_2013,jager_optimal_2014}. These techniques may consist of shaking or moving the trap, or using magnetic fields to modulate the interaction strength via Feshbach resonances.  

In this article we present numerically obtained stationary solutions of the Gross--Pitaevskii (GP) equation in the above described axially symmetric GOST environment. We also numerically estimate the life time of these solutions.

The GP equation describes a system of $N$ particles with local self interaction at zero temperature. This equation can be obtained via the functional derivative 
%\begin{align}
$\ds \ci \hbar \partial_t \Psi = \frac{\delta E \left[ \Psi\right]}{\delta \Psi^*} \equiv H \Psi$ %\label{sec:model:eq1}\end{align}
with respect to the complex conjugate order parameter $\Psi^*$ of the energy functional 
\begin{align}
E \left[ \Psi\right] = \int_{\Omega} \, \left( \frac{\hbar^2}{2 m} \vert \nabla\Psi \vert^2 + V_{\rm ext} \vert\Psi\vert^2 + \frac{g_S}{2} \vert\Psi\vert^4 \right) dV,  \label{sec:model:Energy}
\end{align}
where $\Om\subset\R^3$. In our model the BEC is subject to a gravito-optical surface trap (GOST) which consists of a harmonic radial potential and the Newtonian gravitational potential together with a infinite high wall at $z=0$. Hence the external potential reads 
\begin{align}
V_{\rm ext}(\rho,z) = 
\begin{cases}
\tfrac{1}{2} m \omega^2 \rho^2 + m g z &\text{if } z > 0 \\
\infty & \text{if } z = 0
\end{cases}
\label{sec:gost:pot}
\end{align}
where $\omega$ is the trapping frequency of the harmonic trap, $g$ is the normal gravitational acceleration on earth's surface. We use cylinder coordinates $(\rho,\varphi,z)$ where $\rho$ is the radial coordinate, and $z=0$ describes the reflecting surface of the trap. This surface can be realized experimentally by means of an evanescent laser wave, whereas the harmonic potential may be realized with magnetic or optical traps. The latter confines the BEC in the radial direction $\rho$, while the gravitational acceleration serves as a trap in the vertical $z$-direction. Thus, $\Om=\R^2\times\R_+$ with the boundary condition $\Psi|_{z=0}=0$, and the GP equation 
becomes 
\begin{align}
\ci \hbar \partial_t \Psi = -\frac{\hbar^2}{2 m} \Delta \Psi + V_{\rm ext} \Psi + g_S \vert\Psi\vert^2 \Psi. \label{sec:model:tgpe}
\end{align}

The first contribution is the kinetic energy, the second comes from coupling to the external potential $V_{\rm ext}$, and the last term is due to local self interaction. $g_S=4\pi\hbar^2 a_S / m$ is the coupling strength and is determined by the s-wave scattering length $a_S$. For $g_S>0$ ($g_S<0$) the interaction is repulsive (attractive). In this work we are interested in the first case, so $g_S$ is always positive. Critical points of $E\left[\Phi\right]$ are solutions of the stationary Gross--Pitaevskii equation, which are all degenerate due to U(1) symmetry.  The ground state is a minimum of $E\left[\Phi\right]$, whereas all other critical points are min-max saddle points \cite{rabinowitz_bifurcation_1977}. 

The energy $E$ and the particle number $N := \int_{\Omega} \vert\Psi\vert^2 dV = \Vert \Psi \Vert_{L_2}^2$ are conserved quantities. 
The  separation ansatz $\Psi=\Phi(\rho,\varphi,z) \exp( -\ci \varepsilon t / \hbar )$ in (\ref{sec:model:tgpe}) leads to the stationary GP equation
\begin{align}
\varepsilon \Phi(\rho,\varphi,z) = \left(-\dfrac{\hbar^2}{2m} \Delta + V(\rho,z) + g_S\vert\Phi(\rho,\varphi,z)\vert^2 \right) \Phi(\rho,\varphi,z). \label{sec:model:sgpe_1}
\end{align} 
In cylindrical coordinates we have  
$\ds \Delta := \frac{\partial^2}{\partial\rho^2} + \frac{1}{\rho} \frac{\partial}{\partial\rho} + \frac{1}{\rho^2} \frac{\partial^2}{\partial \varphi^2} + \frac{\partial^2}{\partial z^2} \text{.}$

For a fixed particle number $N$ this equation possesses infinitely many different solutions or, equivalently, infinitely many critical points \cite{rabinowitz_bifurcation_1977}. On unbounded domains $\Omega$ and for potentials which are bounded from below and diverging at infinity, this means that the spectrum consists of discrete eigenvalues $\varepsilon_i\,(i\in\mathbb{N})$, for a fixed particle number. The solution corresponding to the smallest chemical potential $\varepsilon$ is the ground state. Other solutions we will denote as "topological coherent modes" \cite{yukalov_nonlinear_2002}. For all solutions the expectation value of the momentum operator is zero, but only for the ground state the momentum distribution is concentrated around zero. In general, topological coherent modes have also momentum components different from zero. 
Note that (\ref{sec:gost:pot}) is bounded by zero, i.e. $V(\rho,z) \geq 0 $ for $ z \geq 0$, and that the chemical potential $\varepsilon$ can only attain positive values for this potential, otherwise no solutions exist.

In order to make (\ref{sec:model:sgpe_1}) dimensionless we introduce a length scale $L$ and the particle number $N$. 
Replacing 
\begin{align}
\rho \rightarrow L\rho \text{, } z \rightarrow Lz \text{, } \Phi \rightarrow \Phi / L^{3/2} \text{.}
\end{align}
we obtain the dimensionless stationary GP equation 
\begin{align}
\left( -\Delta + \nu^2\rho^2 + \beta z + \gamma \vert\Phi\vert^2 \right) \Phi = \mu \Phi \label{eq:gpe_dimless}
\end{align}
with the dimensionless parameters 
%\allowdisplaybreaks
\begin{align}
\text{trapping frequency } \nu &:= \frac{m \omega}{\hbar} L^2 \text{,} \label{eq:dimless:nu}\\
\text{gravitational acceleration } \beta &:= \frac{2m^2 g}{\hbar^2} L^3 \text{,} \label{eq:dimless:beta}\\
\text{interaction strength } \gamma &:= \frac{2 m g_S}{\hbar^2 L} = 8\pi a_S / L \text{,} \label{eq:dimless:gamma}\\
\text{chemical potential } \mu &:= \frac{2m\varepsilon}{\hbar^2} L^2 \text{,}
\end{align}
which are dependent on the physical parameters and the length scale $L$. Henceforth we use dimensionless quantities and equations until stated otherwise. 

In the following we restrict to pure harmonics in angular direction, i.e., 
we make the ansatz 
\begin{align}
\Phi( \rho, \varphi, z ) = \psilabel(\rho,z) \exp\left( \pm \ci s \varphi \right), \quad \psilabel(\rho,z) \in \mathbb{C},\quad s=0,1,\ldots\, , 
\end{align}
where $\mathfrak{L}$ is used to label different solutions of (\ref{eq:gpe_dimless}). This leads to a simplified Gross-Piatevskii equation  
\begin{align}
\left( H_0 + \gamma \vert \psilabel \vert^2 \right) \psilabel = \mu \psilabel, 
\quad % \label{eq:redgpe}, \end{align}\begin{align}
H_0 := -\frac{\partial^2}{\partial\rho^2} - \frac{1}{\rho} \frac{\partial}{\partial\rho} - \frac{\partial^2}{\partial z^2} + V_{\rm eff}(\rho,z), \label{eq:H0}
\end{align}
with the effective potential
\begin{align}
V_{\rm eff}(\rho,z) := \nu^2\rho^2 + \frac{s^2}{\rho^2} + \beta z, \label{eq:gost_pot_eff}
\end{align}
which contains the centrifugal potential $s^2/\rho^2$, which diverges at $\rho=0$, so that we can expect for $s>0$ vortices at $\rho=0$. The reader may note that $\psilabel$ is not normalized to one.

Thus, our aim is to discuss solutions of \reff{eq:H0} with respect to their stability, and to give estimates of their life time in case of instability. 
In particular we find that many of the unstable solutions decay 
sufficiently slow, so that they might be experimentally observable.  

\section{Stability and life time estimates} 
An overview of techniques to discuss stability in 
Schrödinger type problems can be found in \cite{vakhitov_stationary_1973,  book_kapitula,soffer_resonances_1999,soffer_selection_2004,soffer_theory_2005,zhou_dynamics_2008}. 

The pertinent notion is orbital stability, defined as follows: 
A time harmonic solution $\psilabel$ is called 
orbitally stable if for all $\eps>0$ there exists a $\del>0$ such that 
for all $\Psi_0$ with 
\begin{equation}
\inf_{\vartheta} \Vert \Psi_0 - \psilabel \exp(\ci \vartheta) \Vert_{X} < \delta,
\end{equation}
in some norm $\Vert\cdot\Vert_X$, we have 
\begin{equation}
\inf_{\vartheta} \Vert \Psi(t,\cdot) - \psilabel \exp(\ci \vartheta) \Vert_{X} < \epsilon
\end{equation}
for all times $t>0$, where $\Psi(\cdot,\cdot)$ is the solution to the initial 
condition $\Psi_0$. Thus, if the initial norm of the difference is small, then it remains small for all times, and the solution $\Psi(t,\cdot)$ stays 
close to the ``group orbit'' $\{\psilabel \exp(\ci \vartheta): \vartheta\in [0,2\pi)\}$. 

Essentially, there are four types of (numerical) approaches to study the stability resp. life times of solutions $\psilabel$  (in a discretized setting). 
\bcen
\item Real time propagation \cite{PhysRevB.43.6760, paul_nonlinear_2005}. 
\item Numerical computation of eigenvalues for the linearization of \reff{eq:gpe_dimless} around $\psilabel$. 
\item Complex scaling methods to compute the spectrum \cite{buchleitner_wavefunctions_1994,schlagheck_complex-scaling_2006}. This is also related to the so called quantum mechanical virial theorem, which can be used as an alternative stability check. 
\item  Search for complex eigenvalues via the imaginary time evolution methods  \cite{rapedius_resonance_2009,rapedius_calculating_2011}. 
\ecen 
Our results in \S\ref{rsec} will be based on methods 1 and 2, including a qualitative comparison, but we also use the virial theorem for independent checks.

\subsection{The linearized operator}
The spectrum of a Schrödinger operator can be divided into two parts,  $\sigma_{\rm ess}(H)$ and $\sigma_{\rm d}(H)$. $\sigma_{\rm ess}(H)$ is the essential part, also known as the continuous part, 
determined by the spectrum of the Hamiltonian $H_0=\ci(-\Delta+V_0)$, 
where $V_0\in\R$ denotes a (possible non-zero) limit for $|x|\ra\infty$ of a bounded potential, 
which gives $\sigma_{\rm ess}(H_0)=\{\ci \mu: \mu\ge V_0\}$. 
$\sigma_{\rm d}(H)$ denotes the set of discrete isolated eigenvalues of $H$. For compact perturbations (e.g., changes of the external potential) $\sigma_{\rm d}(H)$ is finite, and the essential spectrum is not altered, due to Weyl's essential spectrum theorem. Embedded eigenvalues can exist inside the essential spectrum which belong to $\sigma_{\rm ess}(H)$ as well. The solutions belonging to such eigenvalues are meta stable and posses an anomalous slow decay, with quantum mechanical tunnelling as the main mechanism for the associated decay of the wave function. 

In order to study the time evolution of a perturbation $h : \Om \times \mathbbm{R} \rightarrow \mathbbm{C}$  we choose the ansatz
\begin{equation}
\psilabel \rightarrow \left( \psilabel + \epsilon h \right) \exp(-\ci \mu t ), 
\end{equation}
where $0 < \epsilon < 1$. This leads to a time dependent GP equation for $h$
\begin{equation}
\ci \partial_t h = \left( H_0 - \mu + 2\gamma\vert\psilabel\vert^2 \right) h + \gamma \psilabel^2 h^* + 2\epsilon\gamma\psilabel \vert h \vert^2 + \epsilon\gamma\psilabel^* h^2 + \gamma \epsilon^2 \vert h \vert^2 h\label{eq:pde_error_h}
\end{equation}
which describes a non unitary time evolution of the perturbation. 
Due to the non-linearity, analytical solutions are hard to obtain, and therefore only the linearized version $\epsilon^0$ is analysed. To order $\epsilon^0$ we obtain by decomposition of $h=h_r+\ci h_i$  and 
\def\psir{\psi_{\mathfrak{L},{\rm r}}}\def\psii{\psi_{\mathfrak{L},{\rm i}}}
$\psilabel=\psir+\ci\psii$ into real and imaginary parts  the 
linear system 
\begin{equation}
\partial_t \begin{pmatrix}
h_r \\
h_i 
\end{pmatrix} = 
\begin{pmatrix}
2\gamma\psir\psii & L^- \\
-L^+ & -2\gamma\psir\psii \\
\end{pmatrix} 
\begin{pmatrix}
h_r \\
h_i 
\end{pmatrix} =: 
M 
\begin{pmatrix}
h_r \\
h_i 
\end{pmatrix}, \label{eq:lin_system_error_h}
\end{equation}
where 
\begin{align}
L^{\pm} &= H_0 - \mu + 2\gamma \vert\psilabel\vert^2 \pm \gamma \left( \psir^2 - \psii^2 \right). %\label{eq:Lminus_op}. 
\label{eq:Lplus_op}
\end{align}
The operator $M$ is not hermitian. Its spectrum consists of $\sigma_{{\rm ess}} = (-\ci\infty,-\ci\mu] \cup [\ci\mu,\ci\infty)$ and  discrete eigenvalues, 
which lie in a strip around the real axis. 
We define a resonance $\mu$ as an eigenvalue of $M$ with $\Im \mu \neq 0$. 
If resonances occur, the life time of $\psilabel$ is defined as 
\begin{equation}
\tau := \frac{\ln 2}{\max\Re{-\ci\sigma_d(M)}}, 
\end{equation}
i.e., $\tau$ is the time required to double the norm $\Vert h \Vert_{L_2}^2$ in the linear evolution.
However, linear stability  (for which formally $\tau=\infty$) does not imply non-linear stability in general, see, e.g., \cite{panos2015}, and a full answer can only be given by studying the full non-linear problem, e.g., using method 1. 

\subsection{The virial theorem} \label{sec:complex_deformations}
The quantum mechanical variant of the virial theorem is 
obtained from the Ehrenfest theorem applied to the von Neumann equation 
\begin{equation}
\frac{d}{dt} \ev{\hat a}= \frac{\ci}{\hbar} \ev{\comm{\hat H}{\hat a}}{\Psi(\cdot,t)} + \ev{\pdv{t} \hat a}{\Psi(\cdot,t)} \text{,} \label{eq:exp_opa}
\end{equation} 
where $\hat{a} := \frac{1}{2} \left( \hat{x}\hat{p} + \hat{p}\hat{x} \right)$ is given in terms of the momentum and position operator $\hat{p}$ and  $\hat{x}$, respectively.
Neither the position operator nor the momentum operator depend explicitly on time, therefore $\pdv{t} \hat{a} = 0$. 
Using the commutator relations 
of the operators in the dimensionless Hamiltonian 
$\hat{H} := \hat{p}^2{+}V({x}){+}\gamma \vert \psilabel({x}) \vert^2$ for the GP equation, i.e., 
\begin{align*}
\ci \commutator{\hat{p}_j^2}{\hat{x}_i} &= -2 \hat{p}_j \delta_{ij},\quad \comm{\hat{p}_i}{\hat{p}_j} = 0, \quad \comm{V(x_i)}{\hat{x}_i}= 0, \\
\ci \comm{\vert\Psi({x}_i,t)\vert^2}{\hat{x}_i} &= 0, \quad \ci \comm{V({x}_i)}{\hat{p}_j} = - \pdv{{x}_i} V(x_i) \delta_{ij} \\ 
\ci \comm{\vert\Psi({x}_i,t)\vert^2}{\hat{p}_j} &= \ci \Psi^*({x}_i,t) \comm{\Psi({x}_i,t)}{\hat{p}_j} + \ci \comm{\Psi^*({x}_i,t)}{\hat{p}_j} \Psi({x}_i,t)\\
&= -\Psi^*({x}_i,t) \pdv{{x}_i,t} \Psi({x}_i,t) \delta_{ij} - \Psi({x}_i,t) \pdv{{x}_i} \Psi^*({x}_i,t) \delta_{ij}, 
\end{align*}
the Ehrenfest Theorem gives 
$%\begin{equation}
\ds \dv{t} \expval{\hat{a}} = -2 \expval{\hat{p}^2}{\Psi} - \expval{{x} \cdot \nabla V({x})}{\Psi} - \gamma\expval{{x} \cdot \nabla \vert\Psi\vert^2}{\Psi}, 
$%\end{equation}
which we write as 
\def\vir{{\rm vir}}
\begin{equation}
\frac{d}{dt} \ev{\hat a} = 2 (T-V) + 3 W =: \vir(t), \label{eq:virial_expression}
\end{equation}
where 
\hualst{
T &:= 2 \pi \int_{\Omega} \vert \nabla \psi(x,t) \vert^2 \,
\rho\,d\rho\,dz\,,\quad 
V := 2 \pi \int_{\Omega} \left( \omega^2 \rho^2 - s^2 \rho^{-2} + \beta z / 2 \right) \vert\psi(x,t)\vert^2 \,\rho\,d\rho\,dz\,,\\
W &:= \pi \gamma \int_{\Omega} \vert\psi(x,t)\vert^4 \,\rho\,d\rho\,dz
} 
are the kinetic, potential, and the self interaction energy, respectively. 
Despite the fact that the term $|\psi|^2\psi$  is not analytic in $\psi$, the contributions of the virial theorem can be formally computed. This is justified by the fact that the absolute square of the wave function can be interpreted as a real potential. If we consider $\ga|\psilabel|^2$ in the GP as a fixed potential, 
then the virial theorem states that $T(t),V(t)$ and $W(t)$ approach constant values as $t\ra\infty$. 
For our real time propagation we also 
compute the quantities $T,V$ and $W$ and in case of stable solutions 
(where $\ga\|\psilabel\|^2$ stays (approximately) fixed) 
these become  constant, whereas for unstable $\psilabel$ they oscillate.

\section{Results}\label{rsec}
We study the stability of ground states and more generally 
the life times of topological coherent modes by solving numerically the Cauchy problem for the time dependent GP equation (\ref{sec:model:tgpe}) with potential (\ref{sec:gost:pot}), and by computing the eigenvalues of the operator $M$ in (\ref{eq:lin_system_error_h}). This has been done several times for different grids, domain sizes, and different time-step lengths. 
The numerical simulations were conducted with our own code \cite{marojevic_atus-pro:_2016} based on the FEM open source library deal.II \cite{dealII84}. The code is written in C++ an can be found at https://github.com/zeli86/atus-pro. For the stationary solutions we used our own Newton method \cite{marojevic_atus-pro:_2016}, and for the time evolution the fully implicit Crank--Nicolson method \cite{numerical_recipes}, which means solving a non-linear set of equations with the standard Newton method. The Crank--Nicolson method is unconditionally stable, and $N$ and $E$ are conserved up to $10^{-8}$ in our simulations, but we obtain a propagating phase error. 
We use Lagrange finite elements  of degree $2$ for each spatial direction. First we used non uniform refined grids with $44071$ cells and $354856$ degrees of freedom with domain sizes $[0,20]\times[0,40]$ and $[0,15]\times[0,30]$. The area of the latter domain corresponds to $[\approx 3.6, \approx 7.3] \mu {\rm m}^2$. The second grid was a regular grid with $65536$ cells and $526338$ degrees of freedom with the same domain size.

We use the error 
\begin{equation}
\kappa(t) := \Vert \Psi(\rho,z,t) - \psi_{\mathfrak{L}}(\rho,z) \exp(-\ci \mu t ) \Vert_{L_2}^2 \label{eq:h_von_t}
\end{equation}
with the $L_2$ norm via the difference of the numerically propagated wave function $\Psi(\rho,z,t)$ and the solutions $\psi_{\mathfrak{L}}(\rho,z) $ of the time independent GP equation (\ref{sec:model:sgpe_1}). This is equivalent to solving the initial value problem (\ref{eq:pde_error_h}). The initial error $\kappa(0)$ is the difference between the numerically computed solution and the true one. This evolution of $\kappa(t)$ is affected by an numerical phase error, however the exponential decay takes place on much shorter time scales than the evolution of the numerical error. Therefore, it is negligible.

As a second criterion we use the first order correlation function 
\begin{equation}
\text{vis}(t) := \left\vert \int_{\Omega} \Psi(\rho,z,t)^* \psi_{\mathfrak{L}}(\rho,z) \rho \, d\rho dz \right\vert /\|\psilabel\|_{L_2}^2
%\left( \int_{\Omega} \psi_{\mathfrak{L}}(\rho,z)^* \psi_{\mathfrak{L}}(\rho,z)  \rho\, d\rho dz \right)^{-1}, 
\label{eq:vis_von_t}
\end{equation}
known as visibility. If there is no visible change in the density of the propagated wave function compared to the density of the initial wave function, then this quantity is equal to one. If the structure of the density starts to dissolve then this function will decrease. Ideally it would drop to zero if all energy is radiated away to infinity, but due to the finite domain and conservation of $N$ this is not possible.

%\subsection{Parameters}
For the physical setup we use $^{87}{\rm Rb}$ with the scattering length $a_S = 90 a_0$ in units of Bohr radii \cite{book_pethick}, which is widely used, e.g., in atom interferometer experiments. For the gravitational trapping we use the earth gravitational acceleration $g = 9.8 \, \rm m \, s^{-2}$, and for the radial trapping we use $\omega = 2\pi$ kHz. For a given length scale of $L = 2.4 \cdot 10^{-7}$ m, which is one order of magnitude bigger than $a_S$, the dimensionless parameter (\ref{eq:dimless:nu})-(\ref{eq:dimless:gamma}) then read 
\begin{align}
\nu &= 0.5, \quad\beta = 0.5, \quad \gamma = 0.5.
\end{align}
The natural time scale is given by $T=2mL^2/\hbar \approx 0.156$ ms, which defines the elementary time unit for our figures. 
The particle number for Rubidium can be computed through
\begin{align}
N_{\rm phys} = \frac{\gamma L}{8\pi a_S} \Vert \psilabel \Vert_{L_2}^2 = 2 \gamma \Vert \psilabel \Vert_{L_2}^2, 
\end{align}
and thus $N_{\rm phys}=\Vert \psilabel \Vert_{L_2}^2$ in our scaling. 

\subsection{Stationary solutions}
Stationary wave functions are obtained numerically via a Newton method \cite{marojevic_atus-pro:_2016} constrained to a special manifold, which allows finding solutions belonging to min-max critical points of the GP functional $E-\mu N$. Alternatively these solutions can be found numerically via  pseudo-arclength continuation, and bifurcation,  \cite{keller_numerical_1977,p2pure, uecker}, see also 
\cite{panos2017} for a recent work displaying a multitude of 
stationary solutions of a GP equation with a parabolic potential. 

Information about the local structure around a solution is provided through the eigenvalues of the second variational derivative of the GP functional (\ref{sec:model:Energy}). If there is a finite number of negative eigenvalues then there is the same finite number of linearly independent descent directions at a critical point. As a consequence, there might be critical points with lower energy so that a part of the energy can decay into these topological coherent modes. 

However if $\mu$ is constant, then the number of solutions is finite. 
The residual of the $L_2$ gradient of our numerically obtained solutions is in the range of $\mathcal{O}(10^{-9})$ to $\mathcal{O}(10^{-10})$. 
This initial residual is considered to be the perturbation for the real time propagation. 

\begin{figure}
\centering
\subfigure[Eight solution branches for (\ref{eq:H0}) in the $\mu$--$N$ plane.]{\scalebox{0.7}{\input{mu_vs_gamma.tex}}}\\
\subfigure[$\vert\psi_{\rm A0}\vert^2$]{\includegraphics[width=35mm]{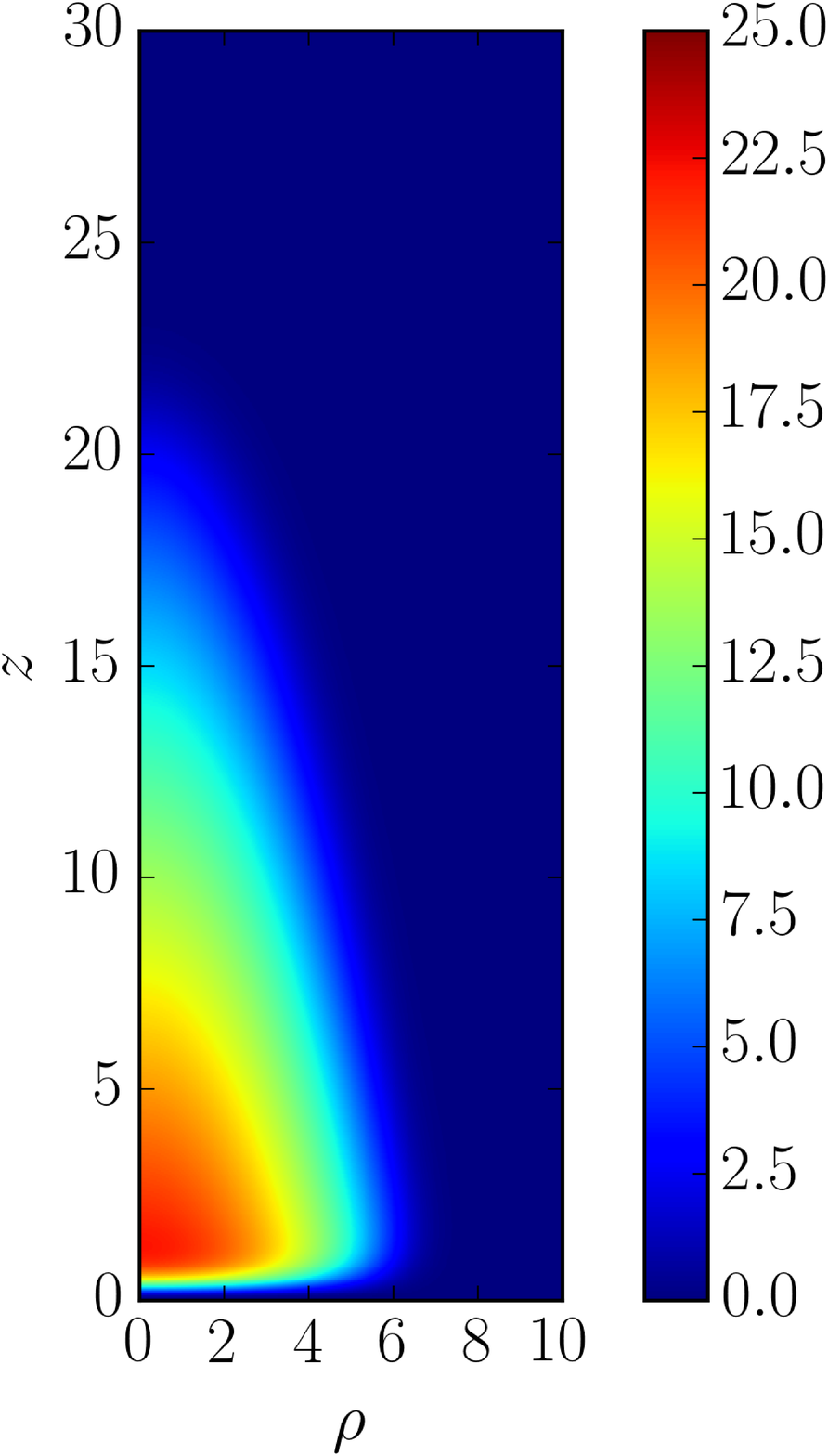}}
\subfigure[$\vert\psi_{\rm B0}\vert^2$]{\includegraphics[width=35mm]{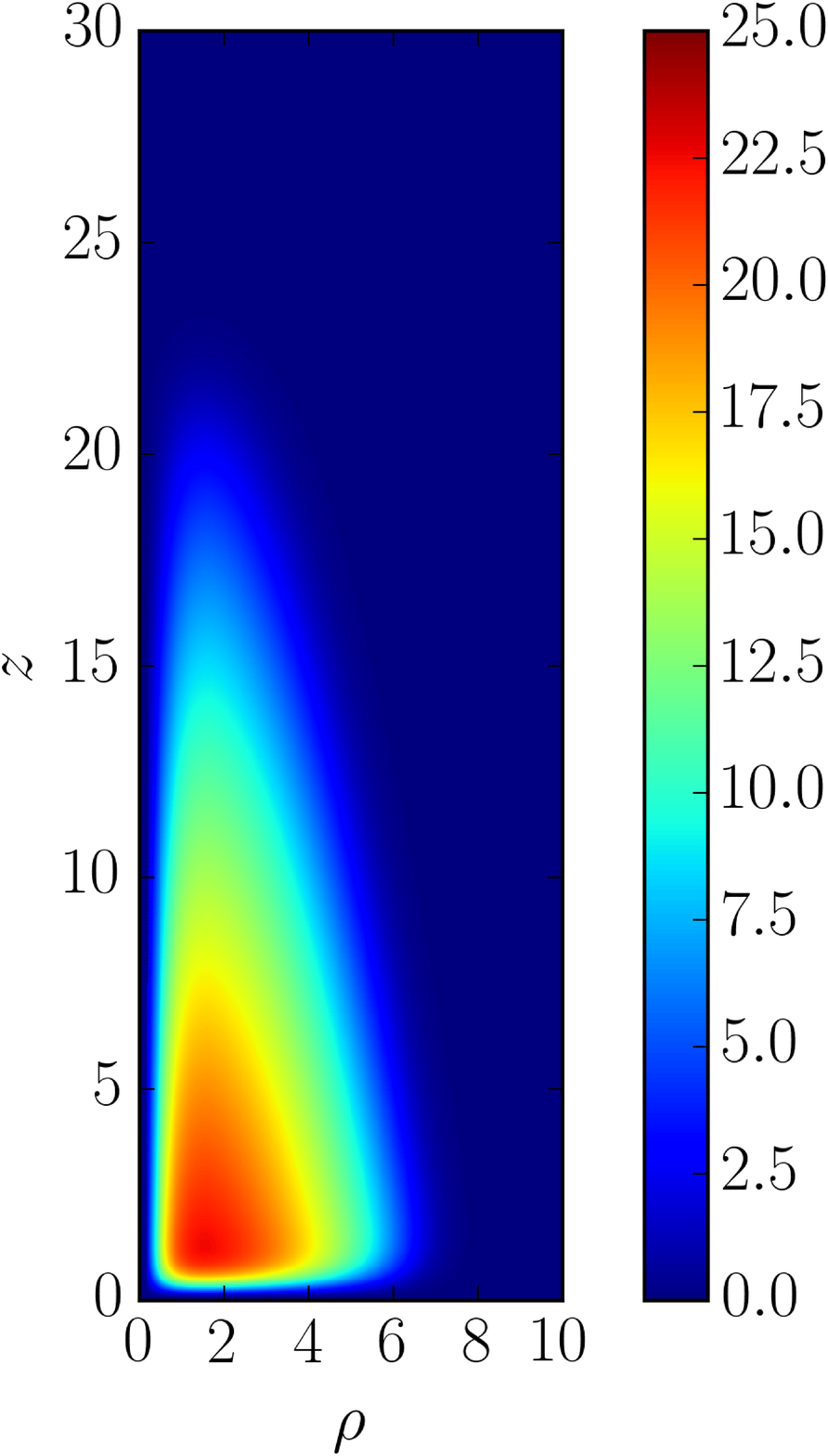}}
\subfigure[$\vert\psi_{\rm A1}\vert^2$]{\includegraphics[width=35mm]{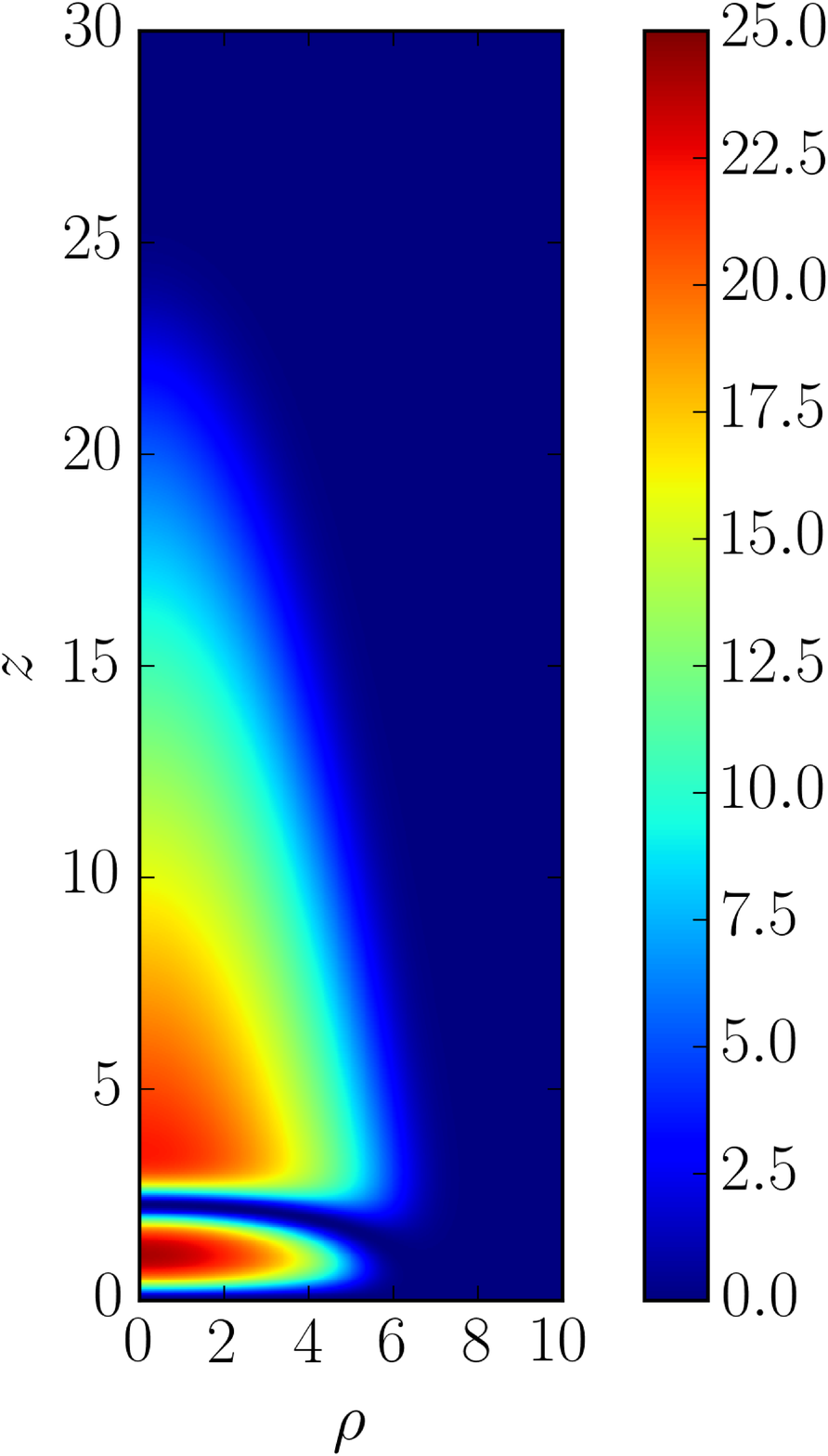}}
\subfigure[$\vert\psi_{\rm A2}\vert^2$]{\includegraphics[width=35mm]{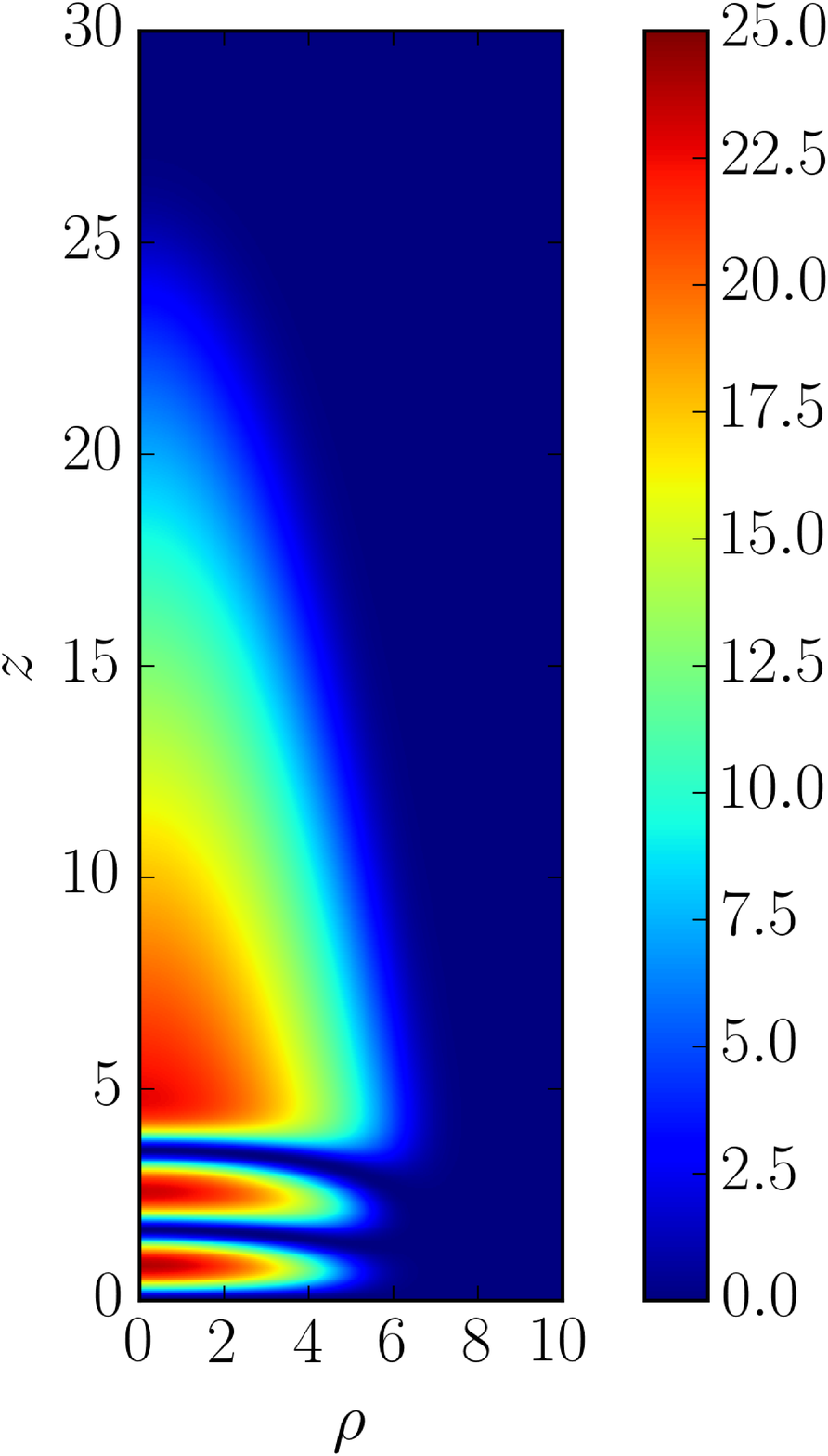}}\\
\subfigure[$\vert\psi_{\rm A3}\vert^2$]{\includegraphics[width=35mm]{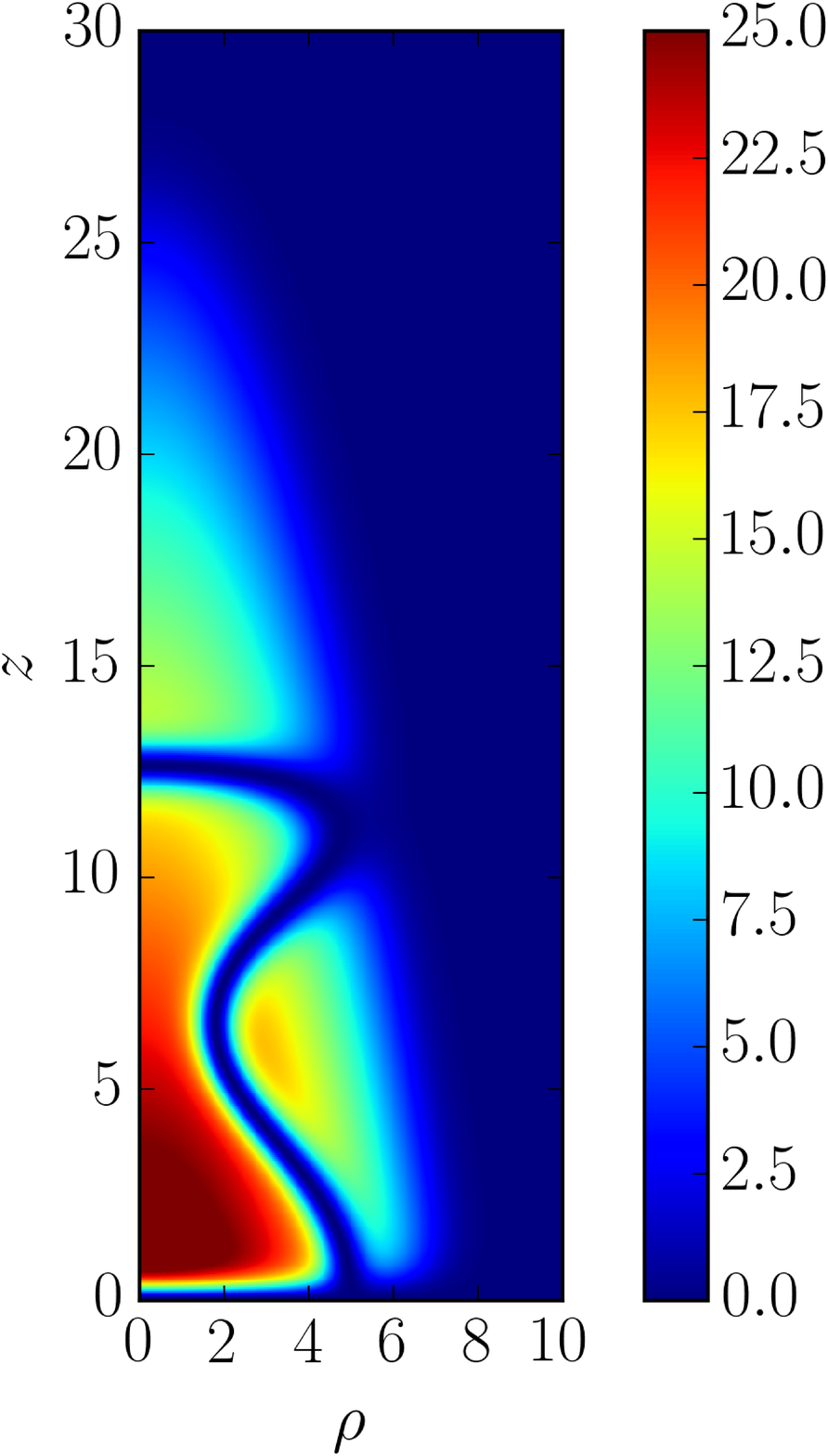}}
\subfigure[$\vert\psi_{\rm B1}\vert^2$]{\includegraphics[width=35mm]{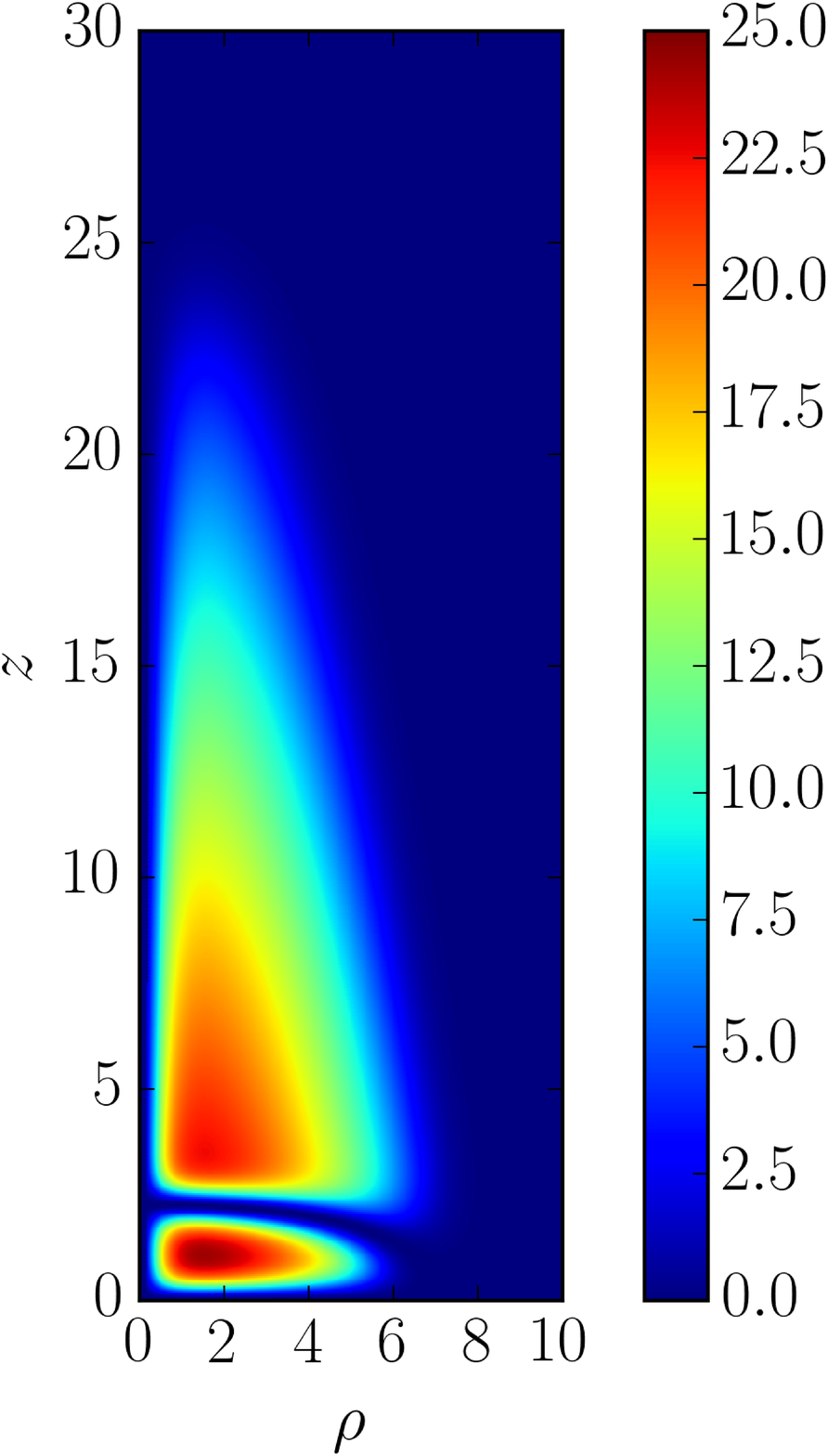}}
\subfigure[$\vert\psi_{\rm B2}\vert^2$]{\includegraphics[width=35mm]{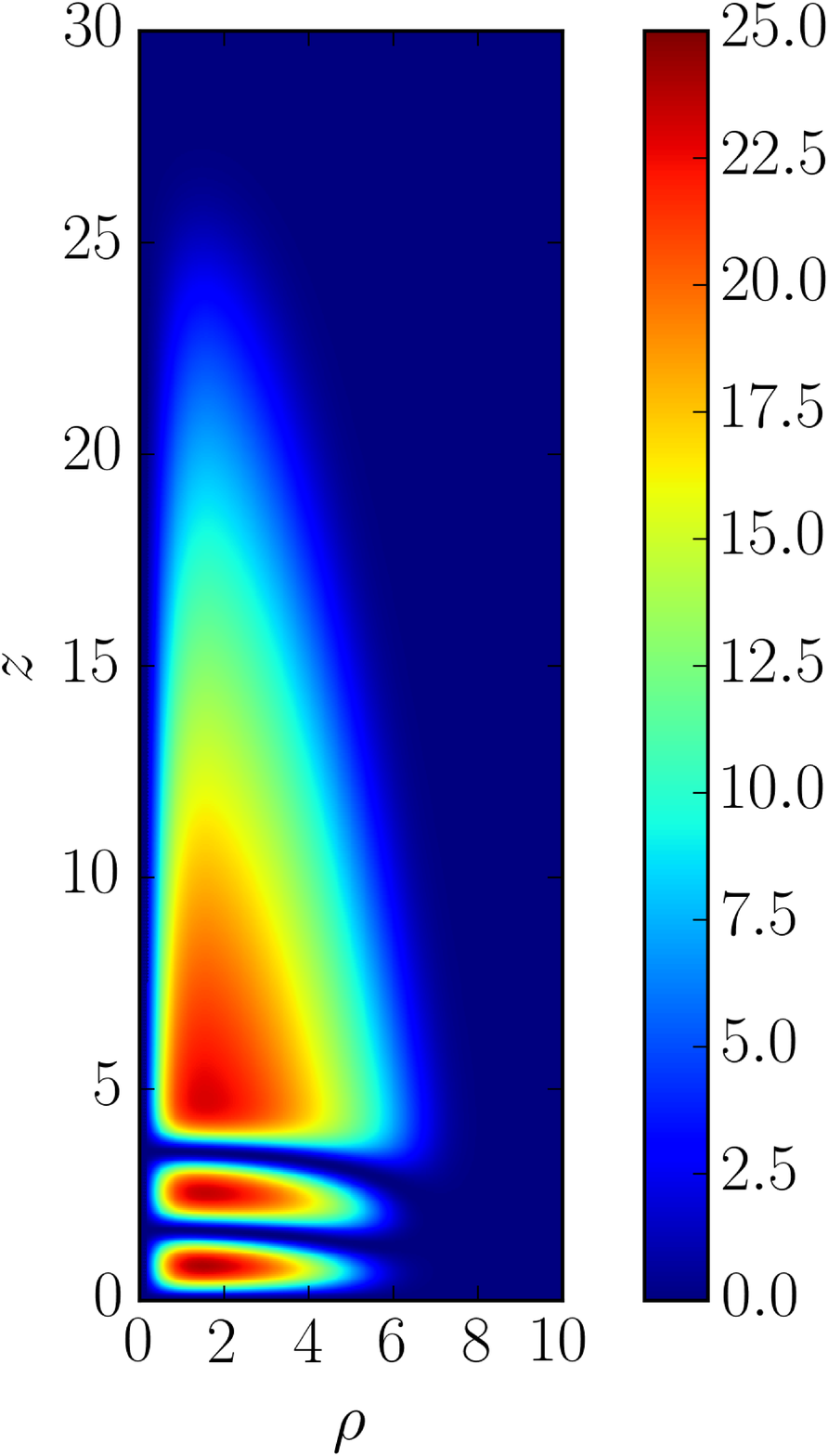}}
\subfigure[$\vert\psi_{\rm B3}\vert^2$]{\includegraphics[width=35mm]{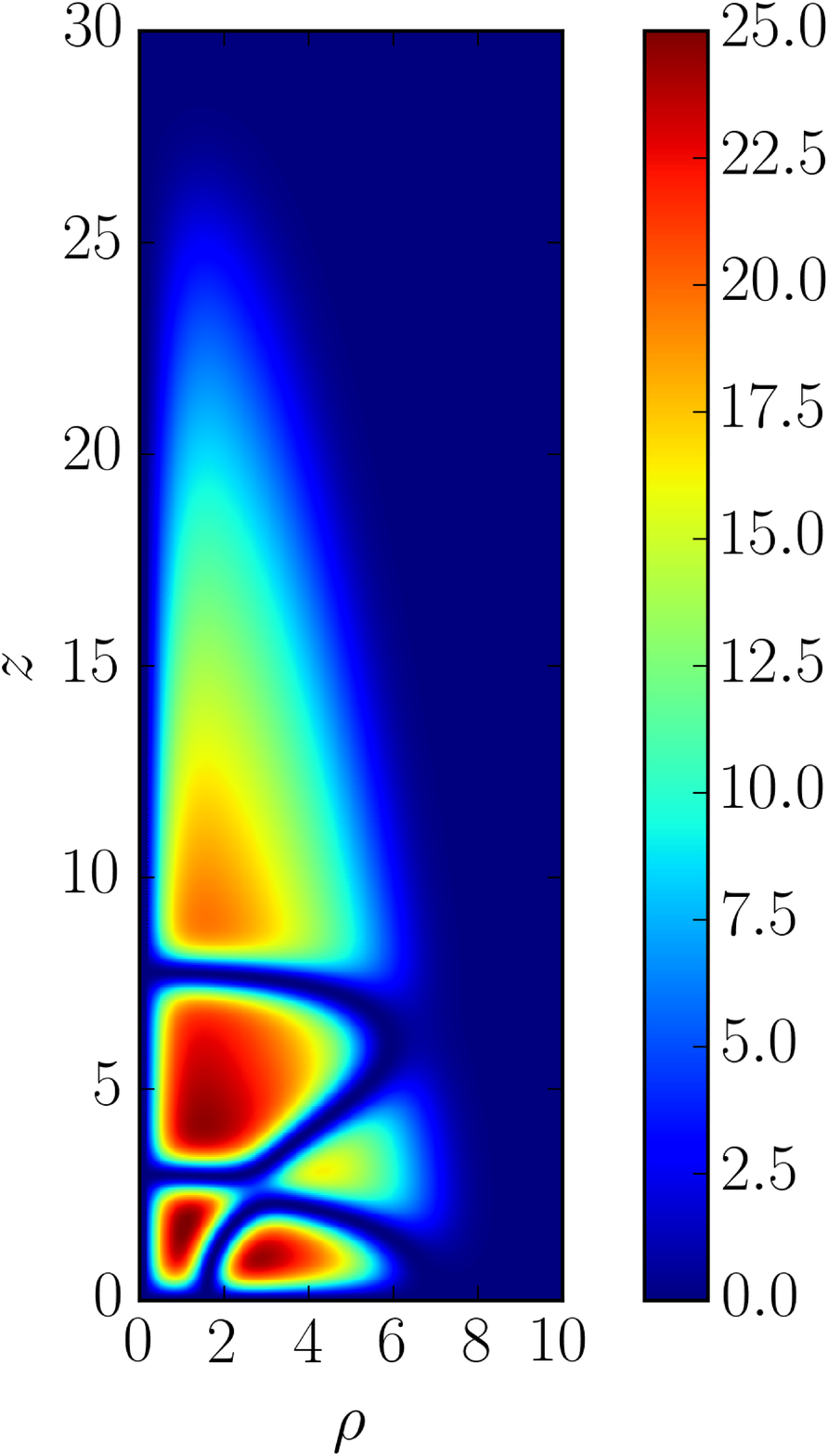}}
\caption{%Example solutions % \protect\linebreak 
Depicted density plots of the eight branches (b)--(i). 
(b) A0, $\mu=11.9$, $N \approx 13058$, $s=0$. %\protect\linebreak 
(c) B0, $\mu=13.1$, $N \approx 16659$, $s=1$. 
(d) A1: $\mu=13$, $N \approx 14776$. % \protect\linebreak 
(e) A2: $\mu=13.9$, $N \approx 16891$. % \protect\linebreak 
(f) A3: $\mu=14.4$, $N \approx 18381$. 
(g) B1: $\mu=14.2$, $N \approx 19298$.  %\protect\linebreak 
(h) B2: $\mu=$, $N \approx 21333$. % \protect\linebreak 
(i) B3: $\mu=15.6$, $N \approx 21892$. 
}\label{fig:gs}
\end{figure}

We have investigated eight solution branches in total, two ground states 
(showing no resonances, i.e., with purely imaginary spectrum), and six topological coherent modes 
(for which we find resonances), see Fig.\ref{fig:gs}.  These eight branches are divided into two different types according to the value of $s$ in (\ref{eq:gost_pot_eff}). The solutions labelled with AM ($M \in \mathbbm{N}$) have $s=0$ and zero angular momentum; solutions labelled with BM have $s=1$ and non-zero angular momentum. For AM we have zero Neumann boundary conditions for $\rho=0$ and zero Dirichlet boundary conditions elsewhere. Concerning BM we have zero Dirichlet boundary conditions on the whole boundary. 

\subsection{Time Evolution}
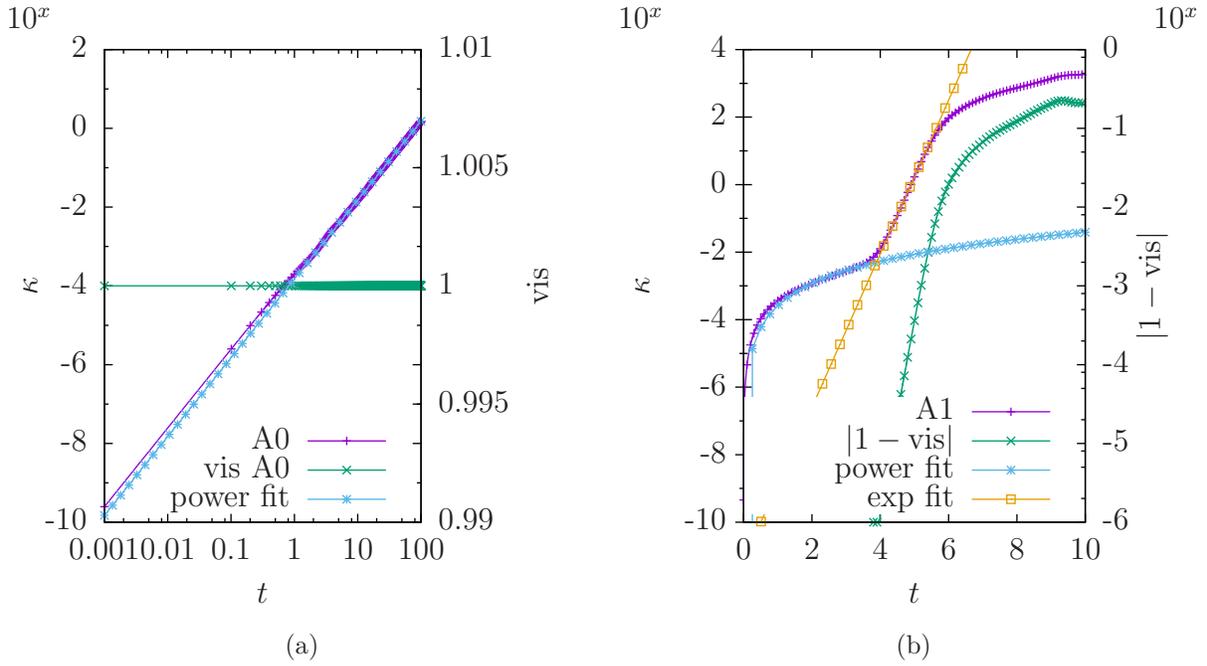
\begin{figure}
\centering
\subfigure[]{\input{err_0_0_0_0090.tex}}
\subfigure[]{\input{err_1_0_0_0090.tex}}
\caption{Examples of $\kappa(t)$ (\ref{eq:h_von_t}) and ${\rm vis}(t)$ (\ref{eq:vis_von_t}) for A0 ($\mu=11.9$,$N=12675$) (a), and A1 ($\mu=13$,$N=14776$) (b).} \label{fig:err_prop_example}
\end{figure}

Figure \ref{fig:err_prop_example}(a) shows the time evolution of the error $\kappa(t)$ (\ref{eq:h_von_t}) (left $y$ axis) and the visibility vis(t) (\ref{eq:vis_von_t}) (right $y$ axis) for the ground state A0 with a large particle number. Although the error $\kappa(t)$ grows due to the propagating phase error, the visibility remains constant which means that there is no change in the structure. The origin of the (unavoidable) phase  error is the discretisation in space and time. We have fitted the error with $\kappa(t)=1.5 \cdot 10^{-4}\,t^{2}$. 
Similar results are obtained for the ground state B0. In summary 
we find that these two ground states are orbitally stable. 

In Fig.~\ref{fig:err_prop_example}(b) a similar plot is depicted for the topological coherent mode A1 with a high particle number, with 
the difference $\vert 1-{\rm vis(t)} \vert$ on the right $x$ axis. 
The time evolution can be separated into three phases. In the time range from $t=0$ to $t\approx 4$ the phase error is dominant, from $t\approx 4$ to $t\approx 5.5$ the non-linearity causes exponential decay, and after $t\approx 5.5$ the structural pattern is fluctuating. In the latter regime the bulk of the density is still confined in the same spatial region due to the trapping potential. In the first phase $\kappa(t)$ grows like $2.62 \cdot 10^{-4}\,t^{2.17}$, and in the second phase like $\exp(5.24\,t-25.67)$. In order to detect the onset of exponential decay the visibility is analysed. It turns out that exponential decay starts when $\vert 1-\text{vis}(t) \vert > 10^{-4}$. For our analysis this defines the lower limit of the time interval with $500$ data points over which we carried out a numerical fit. 

\begin{figure}
\centering
\subfigure[$t=8.5$, vis$\approx 0.993$]{\includegraphics[width=4cm]{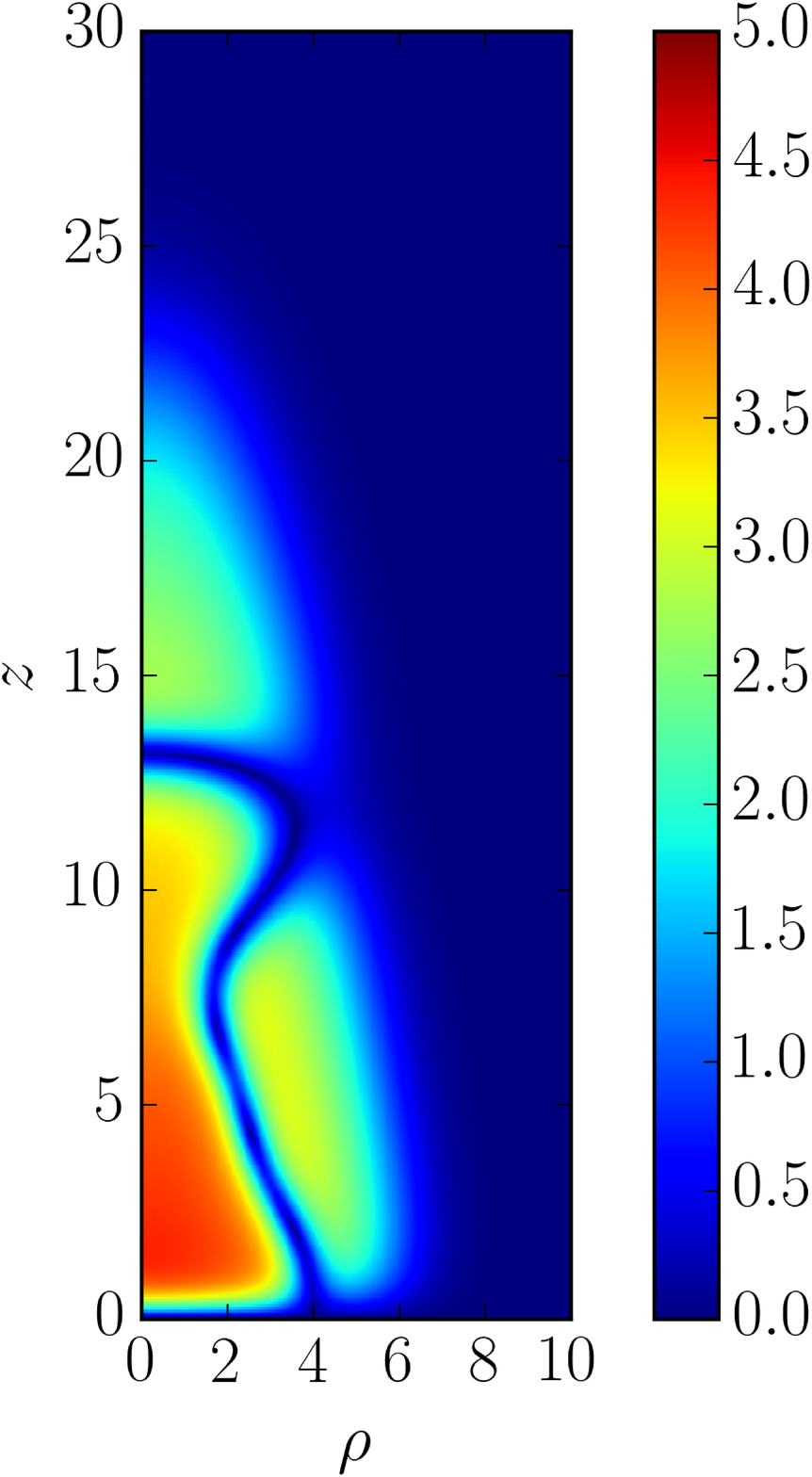}}
\subfigure[$t=9.7$, vis$\approx 0.878$]{\includegraphics[width=4cm]{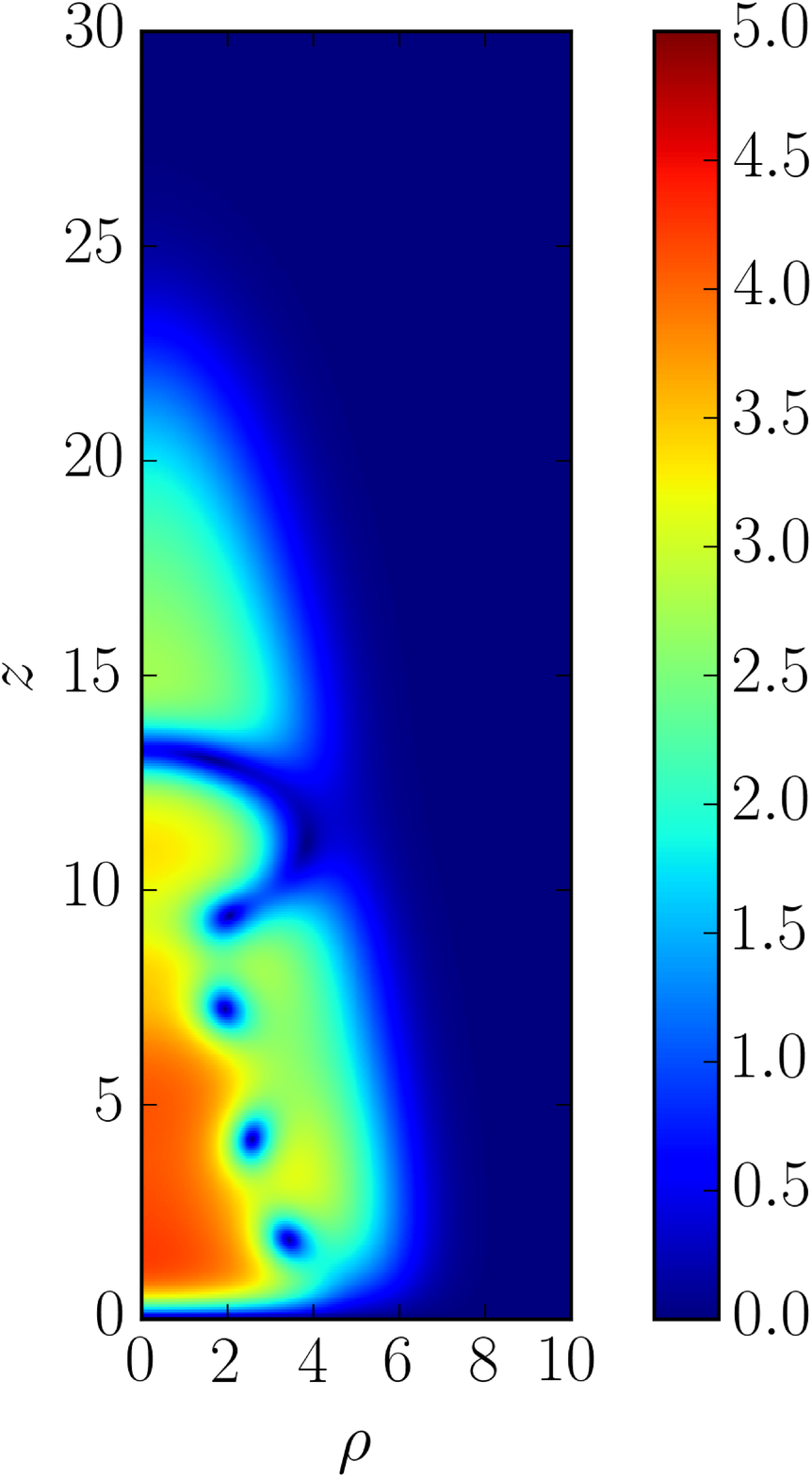}}
\subfigure[$t=11$, vis$\approx 0.779$]{\includegraphics[width=4cm]{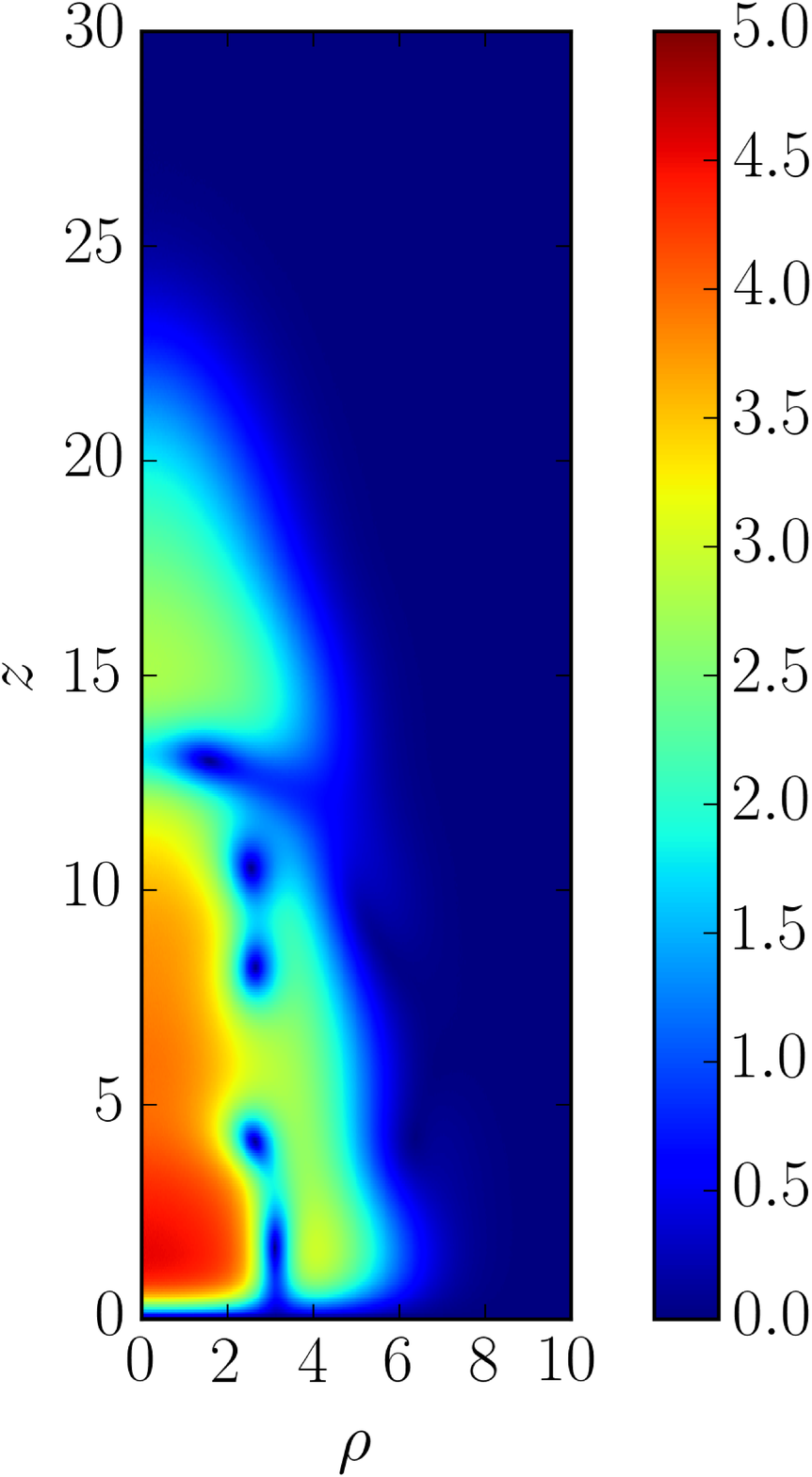}}
\subfigure[$t=20$, vis$\approx 0.446$]{\includegraphics[width=4cm]{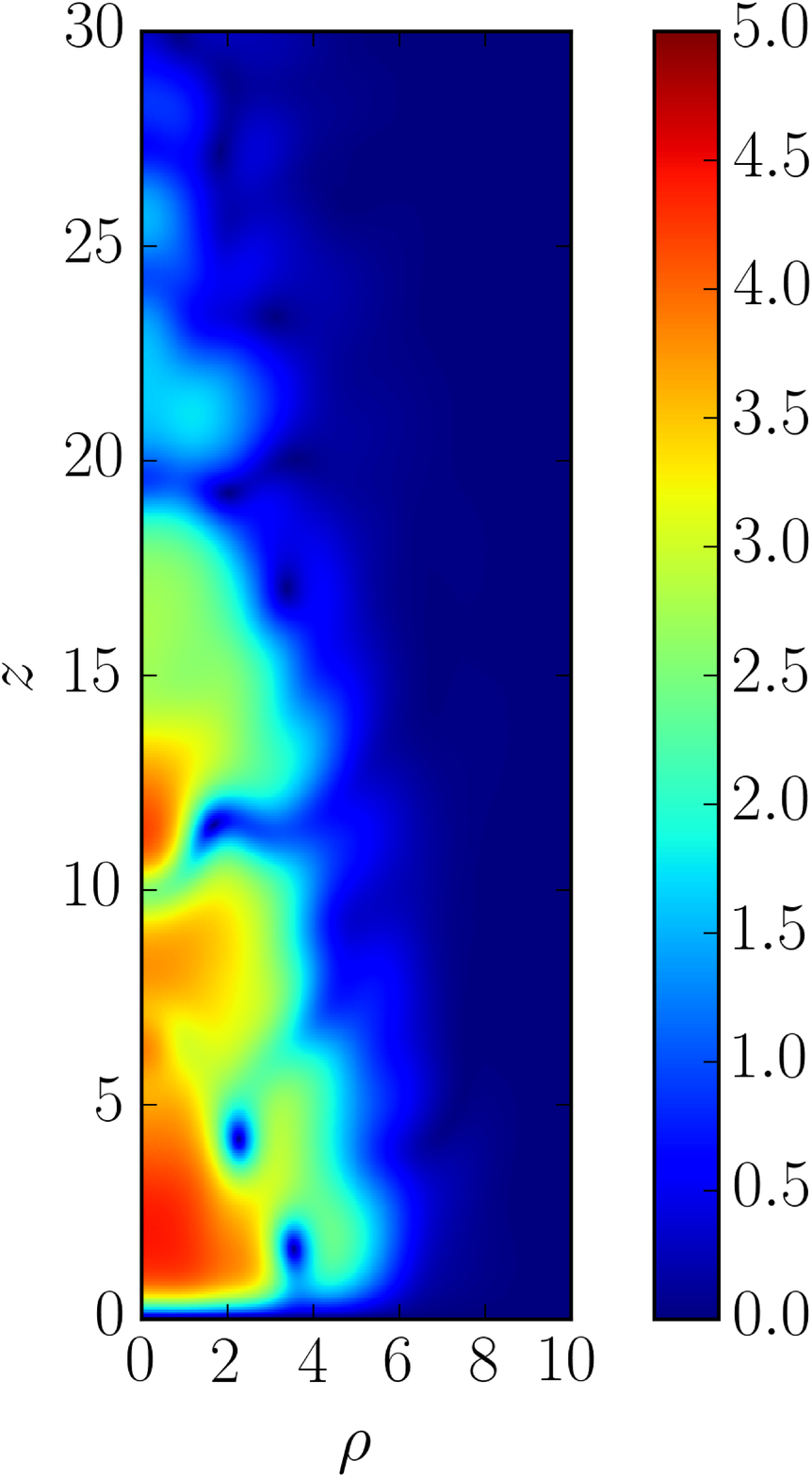}}
\caption{Snapshots from time evolution of the density $\vert \Psi(\rho,z,t) \vert^2$ at different times for the topological coherent mode $A3$, $\mu=10.4$, $N\approx 5598.2$.} \label{fig:snapshots}
\end{figure}

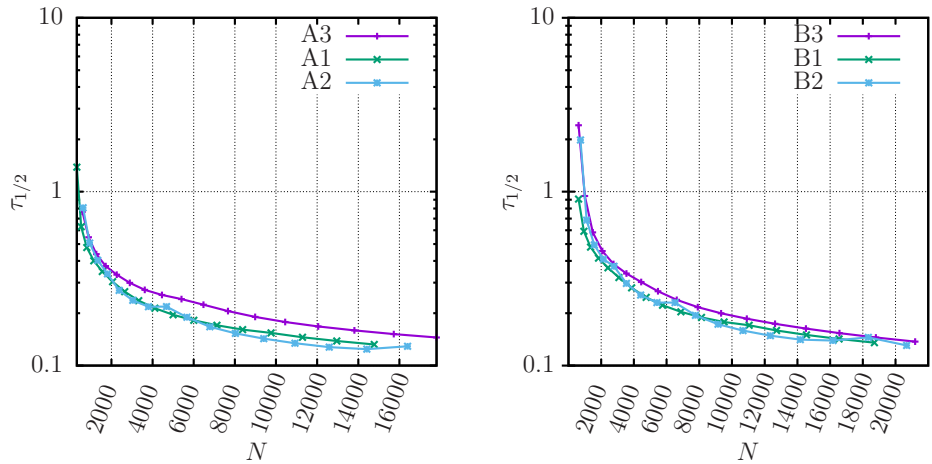
\begin{figure}
\centering
%\subfigure[]
{\scalebox{0.8}{\input{decay_0.tex}}}
%\subfigure[]
{\scalebox{0.8}{\input{decay_1.tex}}}
%\subfigure[]{\scalebox{0.8}{\input{decay_0_cmp_different_omega.tex}}}
%\subfigure[]{\scalebox{0.8}{\input{decay_1_cmp_different_omega.tex}}}
\caption{Half-life estimates for various solutions, depending on $N$. \label{hlfig}}
\end{figure}

In Fig.~\ref{fig:snapshots} an exemplary series of density snapshots of $A3$ at different times is depicted. 
Although in (a) the onset of the exponential decay has already been passed by our definition $\vert 1 - {\rm vis(t)}\vert > 10^{-4}$, the initial structure is still recognizable. The decay starts with a small deformation of the surface of the inner bulk region, which starts to oscillate in time with increasing amplitude until it connects to the right part of the lower bulk of the density, which is visible in (b) and (c). In (d) the structure is strongly dissolved.  
Finally, Fig.~\ref{hlfig} gives $\tau$ for the six unstable solutions which lie within a range of $10^{-5}$ to $10^{-4}$ seconds. 

\subsection{Validation of the results}
For $\gamma=0$ the GP equation (\ref{eq:gpe_dimless}) reduces to a linear Schrödinger equation and the solutions are given by \cite{pauli_vol5}
\begin{equation}
\Phi(\rho,\varphi,z)_{s,k,l} = \sqrt{\dfrac{\nu s!}{\pi(s+k)!}} \exp\left( -\dfrac{\nu \rho^2}{2}\right) \nu^{s/2} \rho^s \text{L}^s_k \left( \nu \rho^2 \right) \exp\left( \pm \ci s \varphi \right) A_l \text{Ai} \left( \beta^{1/3} z + z_l \right), \label{eq:linsols}
\end{equation}
where $s=0,1,2,\ldots $ and $k=0,1,2,\ldots $ are the angular and radial momentum quantum numbers, respectively, ${\rm L}_k^s$ are the Laguerre polynomials 
$\ds %\begin{align}
\text{L}^s_k (x) := \sum_{i=0}^{k} (-1)^i \frac{(s+k)!}{(k-i)!(s+i)! i!}  x^i \text{,}
$ %\end{align}
 and $l=0,1,\ldots $ is the quantum number belonging to the gravitational trapping, 
with the $z_l$ given by the zeroes of the Airy Ai function, which is normalized by
$%\begin{align}
\ds A_l := \left( \int_0^\infty \, \text{Ai} \left( \beta^{1/3} z - \vert z_l \vert \right)^2 dz \right)^{-1/2} \text{.}
$ %end{align}
The spectrum is given by
\begin{align}
\varepsilon_{s,k,l} = \nu \left( s + 2k + 1 \right) + \beta^{2/3}  \vert z_l \vert \label{eq:lin_spectrum}
\end{align}
where the eigenstates $\Psi_{s,k,l}$ are $(s+2k+1)$ fold degenerated.
The eigenfunctions (\ref{eq:linsols}) span a complete orthonormal basis of 
$L^2(\Om)$, $\Om=\R^2\times\R_+$, with respect to the inner product 
\begin{equation}
\int_{0}^{\infty} \int_{0}^{2\pi} \int_{0}^{\infty} \Phi(\rho,\varphi,z)^*_{s,k,l} \Phi(\rho,\varphi,z)_{s^{\prime},k^{\prime},l^{\prime}} \, \rho \, d\rho  \, d\varphi \, dz = \delta_{s,s^{\prime}} \delta_{k,k^{\prime}} \delta_{l,l^{\prime}}.
\end{equation}

In order to check the quality of the half life time estimates obtained via the real time propagation we compared them to the eigenvalues of $M$ in (\ref{eq:lin_system_error_h}), obtained from expanding $M$ in 
the basis (\ref{eq:linsols}). 
The eigenvalue with the largest real part is then used for the estimate. 
The operator $L^-$ (see (\ref{eq:Lplus_op})) expanded in the basis (\ref{eq:linsols}) reads  
\begin{align}
M_{N^2+j+i N, k+l N} &= \int \nabla \Phi_{s,i,j} \nabla\Phi_{s,k,l} + (V_{\rm ext} - \mu + \gamma \psilabel^2) \Phi_{s,i,j} \Phi_{s,k,l} \, \rho \, d\rho d\varphi dz \notag\\
 &= \left( \nu \left( s + 2k + 1 \right) + \beta^{2/3} \vert z_l \vert - \mu \right) \delta_{i,k} \delta_{j,l} + \int \gamma \psilabel^2 \Phi_{s,i,j} \Phi_{s,k,l} \, \rho \, d\rho d\varphi dz,
\end{align} 
where $N$ is the highest quantum number for $i$ and $j$, respectively, $k$ and $l$, and $s$ is fixed. The dimension of the basis is thus $N^2$, 
where we used   $N{=}400$. This corresponds to the upper right block of $M$. 

The lower left block is expanded analogously. The remaining entries of $M$ are zero because $\Im{\psilabel}{=}0$. Expansions of this type are often much more 
efficient than computing eigenvalues directly from the (large) Jacobian  matrix of the numerical solution in the FEM setting \cite{book_yang}. 

\begin{figure}
\centering
\subfigure[]{\includegraphics[width=8cm]{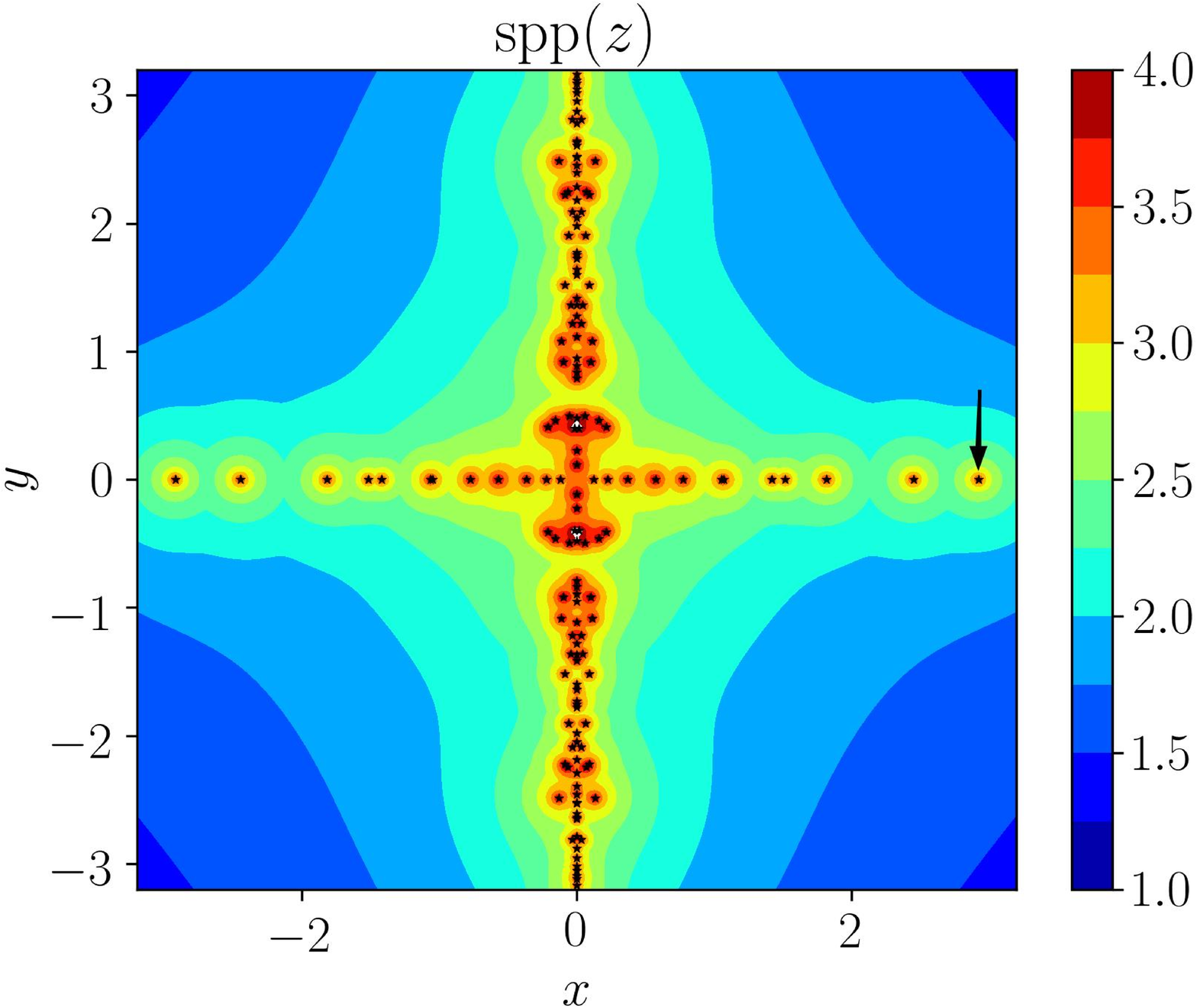}} 
\subfigure[]{\input{cross_section_demo_010_0050.tex}}
\caption{Spectral portrait of solution A3 with $\mu=10.4$ and $N\approx 5405$. (a) Section in the complex plane, where $z=x+\ci y$. The black dots shows the position of the eigenvalues. The arrow indicates the eigenvalue used for the estimate. (b) Cross section along the real axis through the marked eigenvalue (right peak). } \label{fig:spp}
\end{figure}

In Fig.~\ref{fig:spp}(a) an exemplary section of the spectral portrait of $M$ is depicted. The spectral portrait 
\begin{equation}
\text{spp}(z) := \log_{10} \left( \Vert \left(zI - M \right)^{-1} \Vert_2 \Vert M \Vert_2 \right), 
\end{equation}
where $\Vert \cdot \Vert_2$ is the usual matrix norm, 
is a useful tool to check how trustworthy numerically computed eigenvalues are. This is important for studying the stability of certain numerical schemes, see for example \cite{trefethen_embree}. The idea behind this is that for a given eigenvalue $z_0$ we have $\lim_{z\to z_0}\Vert \left(z_0 I - M \right)^{-1} \Vert=\infty$.  
Thus, if the matrix $M$ is perturbed or known by a relative error up to $\epsilon$, then the numerically computed eigenvalue has an uncertainty enclosed by the region where $\text{spp}(z) > \epsilon^{-1}$. In Fig.~\ref{fig:spp}(b) a cross section of (a) along the real axes through the marked eigenvalue (arrow) is shown. 

\begin{figure}[htbp]
\centering
\scalebox{0.8}{\input{decay_A1_cmp.tex}}
\scalebox{0.8}{\input{decay_B1_cmp.tex}}
\scalebox{0.8}{\input{decay_A3_cmp.tex}}
\scalebox{0.8}{\input{decay_B3_cmp.tex}}
% \subfigure[A1]{\scalebox{0.8}{\input{decay_A1_cmp.tex}}}
% \subfigure[B1]{\scalebox{0.8}{\input{decay_B1_cmp.tex}}}
% \subfigure[A3]{\scalebox{0.8}{\input{decay_A3_cmp.tex}}}
% \subfigure[B3]{\scalebox{0.8}{\input{decay_B3_cmp.tex}}}
\caption{The error bars indicates the size of the region where $\text{spp}(z)>10^5$.} \label{fig:A1_B1_comp}
\end{figure}
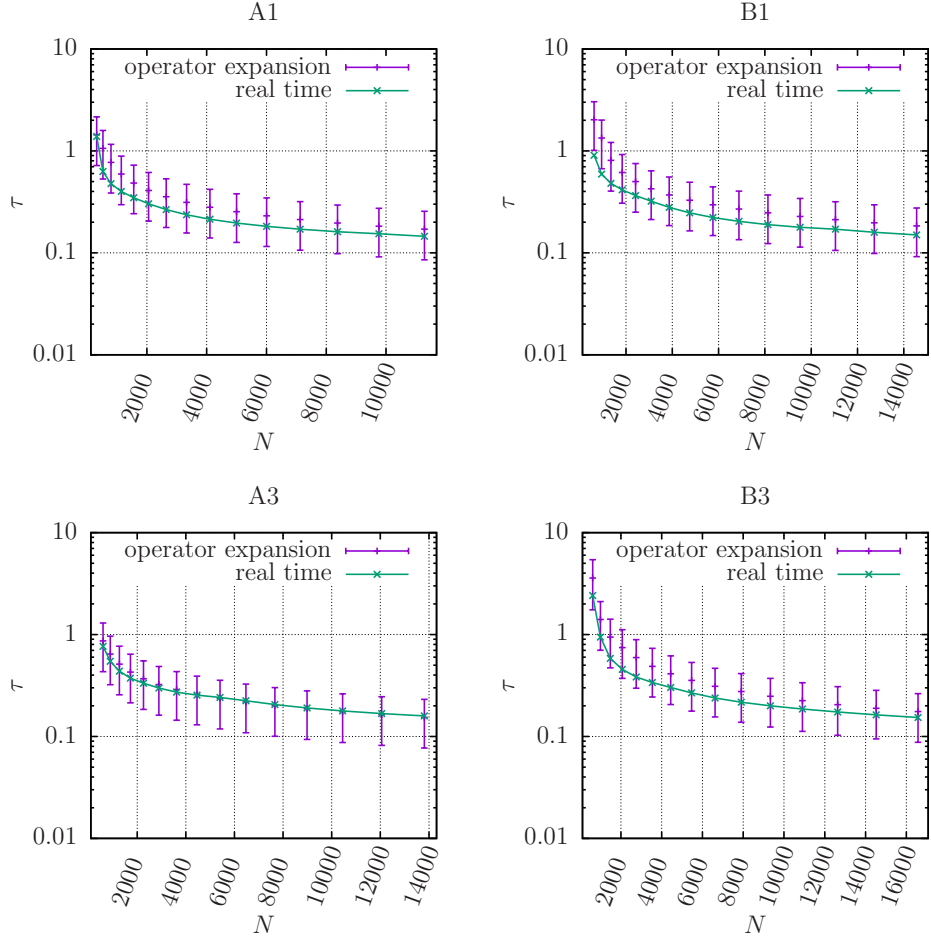

Figure \ref{fig:A1_B1_comp} 
displays the comparison between both methods. 
The error bars are computed from the spectral portrait for $\epsilon^{-1} = 10^{-5}$, which is the smallest $\epsilon$ with no significant change in the size of the error bars.

As a third method to assess the stability of stationary states we may use 
\reff{eq:virial_expression}, implying that $\vir(t):=2 (T(t)-V(t)) + 3 W(t)\approx 0$ 
as long as we stay close to a (time-harmonic) $\psilabel$. In the numerics 
we find that indeed $\vir(t)$ stays very close to $0$ for stable states, 
while $\vir(t)$ starts to oscillate once the instability of an unstable 
state begins to manifest. 

\section{Summary and Outlook} 
We presented numerical life time estimates for a selection of stationary solutions for the GP equation with cylindrical symmetry for a gravito optical surface trap (GOST). For all coherent topological states the numerical life time estimates lie in the range of $10^{-4}$ to $10^{-5}$ seconds. This would be accessible by experiments, if realizable. 
One way for assisting experimental realization could be to compute translations and deformations of a time dependent external potential by quantum optimal control techniques \cite{bucker_twin-atom_2011,bucker_vibrational_2013}. By means of this transitions from ground states to desired topological coherent states can be achieved. After preparing a topological coherent state in a GOST, the life time could by determined through experiments and compared to our  numerical estimates. The next step could be to 
release such states from the trap and to try to capture their structure during a free propagation, for example under weightlessness conditions. 

Another application could be a test of the equivalence principle. Here, no difference between gravitational and inertial mass has been made. Therefore, if the ratio of inertial and gravitational mass should differ for different atomic species this could lead to a deviation of solution branches compared to the solution branches presented in this work. 
This could also help to detect inconsistencies. 

\section*{Acknowledgement}
We like to thank A. Balaz, J. Kunz, A. Pelster for fruitful discussions and T. S. Lee for giving suggestions for improvements of the manuscript. This research is supported by the German Space Agency (DLR) with funds provided by the Federal Ministry for Economic Affairs and Energy (BMWi) due to an enactment of the German Bundestag under grant number 50WM1642. Furthermore, use of the HPC resources of the Nort-German Supercomputing Alliance (HLRN) is gratefully acknowledged. Finally, we acknowledge support from the DFG within the Research Training Group 1620 ``Models of Gravity''. 

%\printbibliography
\bibliographystyle{elsarticle-num}
\bibliography{paper}

\end{document}

%% file: hudef-b.tex
\newcommand{\reff}[1]{(\ref{#1})}
\newcommand{\bga}{\begin{gather}}\newcommand{\ega}{\end{gather}}
\newcommand{\bi}{\begin{itemize}}\newcommand{\ei}{\end{itemize}}
\newcommand{\ba}{\begin{array}}\newcommand{\ea}{\end{array}}
\newcommand{\barr}{\begin{array}}\newcommand{\earr}{\end{array}}
\newcommand{\btab}{\begin{tabular}}\newcommand{\etab}{\end{tabular}}
\newcommand{\bpm}{\begin{pmatrix}}\newcommand{\epm}{\end{pmatrix}}

\def\ga{\gamma}

\def\R{\mathbb{R}}

\def\eps{\varepsilon}
\def\ra{\rightarrow}

\def\ga{\gamma}\def\del{\delta}

%% file: mu_vs_gamma.tex
% GNUPLOT: LaTeX picture with Postscript
\begingroup
  \makeatletter
  \providecommand\color[2][]{%
    \GenericError{(gnuplot) \space\space\space\@spaces}{%
      Package color not loaded in conjunction with
      terminal option `colourtext'%
    }{See the gnuplot documentation for explanation.%
    }{Either use 'blacktext' in gnuplot or load the package
      color.sty in LaTeX.}%
    \renewcommand\color[2][]{}%
  }%
  \providecommand\includegraphics[2][]{%
    \GenericError{(gnuplot) \space\space\space\@spaces}{%
      Package graphicx or graphics not loaded%
    }{See the gnuplot documentation for explanation.%
    }{The gnuplot epslatex terminal needs graphicx.sty or graphics.sty.}%
    \renewcommand\includegraphics[2][]{}%
  }%
  \providecommand\rotatebox[2]{#2}%
  \@ifundefined{ifGPcolor}{%
    \newif\ifGPcolor
    \GPcolortrue
  }{}%
  \@ifundefined{ifGPblacktext}{%
    \newif\ifGPblacktext
    \GPblacktexttrue
  }{}%
  % define a \g@addto@macro without @ in the name:
  \let\gplgaddtomacro\g@addto@macro
  % define empty templates for all commands taking text:
  \gdef\gplbacktext{}%
  \gdef\gplfronttext{}%
  \makeatother
  \ifGPblacktext
    % no textcolor at all
    \def\colorrgb#1{}%
    \def\colorgray#1{}%
  \else
    % gray or color?
    \ifGPcolor
      \def\colorrgb#1{\color[rgb]{#1}}%
      \def\colorgray#1{\color[gray]{#1}}%
      \expandafter\def\csname LTw\endcsname{\color{white}}%
      \expandafter\def\csname LTb\endcsname{\color{black}}%
      \expandafter\def\csname LTa\endcsname{\color{black}}%
      \expandafter\def\csname LT0\endcsname{\color[rgb]{1,0,0}}%
      \expandafter\def\csname LT1\endcsname{\color[rgb]{0,1,0}}%
      \expandafter\def\csname LT2\endcsname{\color[rgb]{0,0,1}}%
      \expandafter\def\csname LT3\endcsname{\color[rgb]{1,0,1}}%
      \expandafter\def\csname LT4\endcsname{\color[rgb]{0,1,1}}%
      \expandafter\def\csname LT5\endcsname{\color[rgb]{1,1,0}}%
      \expandafter\def\csname LT6\endcsname{\color[rgb]{0,0,0}}%
      \expandafter\def\csname LT7\endcsname{\color[rgb]{1,0.3,0}}%
      \expandafter\def\csname LT8\endcsname{\color[rgb]{0.5,0.5,0.5}}%
    \else
      % gray
      \def\colorrgb#1{\color{black}}%
      \def\colorgray#1{\color[gray]{#1}}%
      \expandafter\def\csname LTw\endcsname{\color{white}}%
      \expandafter\def\csname LTb\endcsname{\color{black}}%
      \expandafter\def\csname LTa\endcsname{\color{black}}%
      \expandafter\def\csname LT0\endcsname{\color{black}}%
      \expandafter\def\csname LT1\endcsname{\color{black}}%
      \expandafter\def\csname LT2\endcsname{\color{black}}%
      \expandafter\def\csname LT3\endcsname{\color{black}}%
      \expandafter\def\csname LT4\endcsname{\color{black}}%
      \expandafter\def\csname LT5\endcsname{\color{black}}%
      \expandafter\def\csname LT6\endcsname{\color{black}}%
      \expandafter\def\csname LT7\endcsname{\color{black}}%
      \expandafter\def\csname LT8\endcsname{\color{black}}%
    \fi
  \fi
    \setlength{\unitlength}{0.0500bp}%
    \ifx\gptboxheight\undefined%
      \newlength{\gptboxheight}%
      \newlength{\gptboxwidth}%
      \newsavebox{\gptboxtext}%
    \fi%
    \setlength{\fboxrule}{0.5pt}%
    \setlength{\fboxsep}{1pt}%
\begin{picture}(9070.00,4534.00)%
    \gplgaddtomacro\gplbacktext{%
      \csname LTb\endcsname%
      \put(946,704){\makebox(0,0)[r]{\strut{}$0$}}%
      \csname LTb\endcsname%
      \put(946,1150){\makebox(0,0)[r]{\strut{}$2000$}}%
      \csname LTb\endcsname%
      \put(946,1595){\makebox(0,0)[r]{\strut{}$4000$}}%
      \csname LTb\endcsname%
      \put(946,2041){\makebox(0,0)[r]{\strut{}$6000$}}%
      \csname LTb\endcsname%
      \put(946,2487){\makebox(0,0)[r]{\strut{}$8000$}}%
      \csname LTb\endcsname%
      \put(946,2932){\makebox(0,0)[r]{\strut{}$10000$}}%
      \csname LTb\endcsname%
      \put(946,3378){\makebox(0,0)[r]{\strut{}$12000$}}%
      \csname LTb\endcsname%
      \put(946,3823){\makebox(0,0)[r]{\strut{}$14000$}}%
      \csname LTb\endcsname%
      \put(946,4269){\makebox(0,0)[r]{\strut{}$16000$}}%
      \csname LTb\endcsname%
      \put(1078,484){\makebox(0,0){\strut{}$4$}}%
      \csname LTb\endcsname%
      \put(1827,484){\makebox(0,0){\strut{}$5$}}%
      \csname LTb\endcsname%
      \put(2575,484){\makebox(0,0){\strut{}$6$}}%
      \csname LTb\endcsname%
      \put(3324,484){\makebox(0,0){\strut{}$7$}}%
      \csname LTb\endcsname%
      \put(4072,484){\makebox(0,0){\strut{}$8$}}%
      \csname LTb\endcsname%
      \put(4821,484){\makebox(0,0){\strut{}$9$}}%
      \csname LTb\endcsname%
      \put(5569,484){\makebox(0,0){\strut{}$10$}}%
      \csname LTb\endcsname%
      \put(6318,484){\makebox(0,0){\strut{}$11$}}%
      \csname LTb\endcsname%
      \put(7066,484){\makebox(0,0){\strut{}$12$}}%
      \csname LTb\endcsname%
      \put(7815,484){\makebox(0,0){\strut{}$13$}}%
      \csname LTb\endcsname%
      \put(8563,484){\makebox(0,0){\strut{}$14$}}%
    }%
    \gplgaddtomacro\gplfronttext{%
      \csname LTb\endcsname%
      \put(176,2486){\rotatebox{-270}{\makebox(0,0){\strut{}$N$}}}%
      \csname LTb\endcsname%
      \put(8782,2486){\rotatebox{-270}{\makebox(0,0){\strut{}}}}%
      \csname LTb\endcsname%
      \put(4820,154){\makebox(0,0){\strut{}$\mu$}}%
      \csname LTb\endcsname%
      \put(4820,4159){\makebox(0,0){\strut{}}}%
      \csname LTb\endcsname%
      \put(4820,4158){\makebox(0,0){\strut{}}}%
      \csname LTb\endcsname%
      \put(286,110){\makebox(0,0)[l]{\strut{}}}%
      \put(1864,4049){\makebox(0,0){\strut{}}}%
      \csname LTb\endcsname%
      \put(2518,4048){\makebox(0,0)[r]{\strut{}A0}}%
      \csname LTb\endcsname%
      \put(2518,3733){\makebox(0,0)[r]{\strut{}B0}}%
      \csname LTb\endcsname%
      \put(2518,3418){\makebox(0,0)[r]{\strut{}A1}}%
      \csname LTb\endcsname%
      \put(2518,3103){\makebox(0,0)[r]{\strut{}B1}}%
      \csname LTb\endcsname%
      \put(2518,2788){\makebox(0,0)[r]{\strut{}A2}}%
      \csname LTb\endcsname%
      \put(2518,2473){\makebox(0,0)[r]{\strut{}A3}}%
      \csname LTb\endcsname%
      \put(2518,2158){\makebox(0,0)[r]{\strut{}B2}}%
      \csname LTb\endcsname%
      \put(2518,1843){\makebox(0,0)[r]{\strut{}B3}}%
    }%
    \gplbacktext
    \put(0,0){\includegraphics{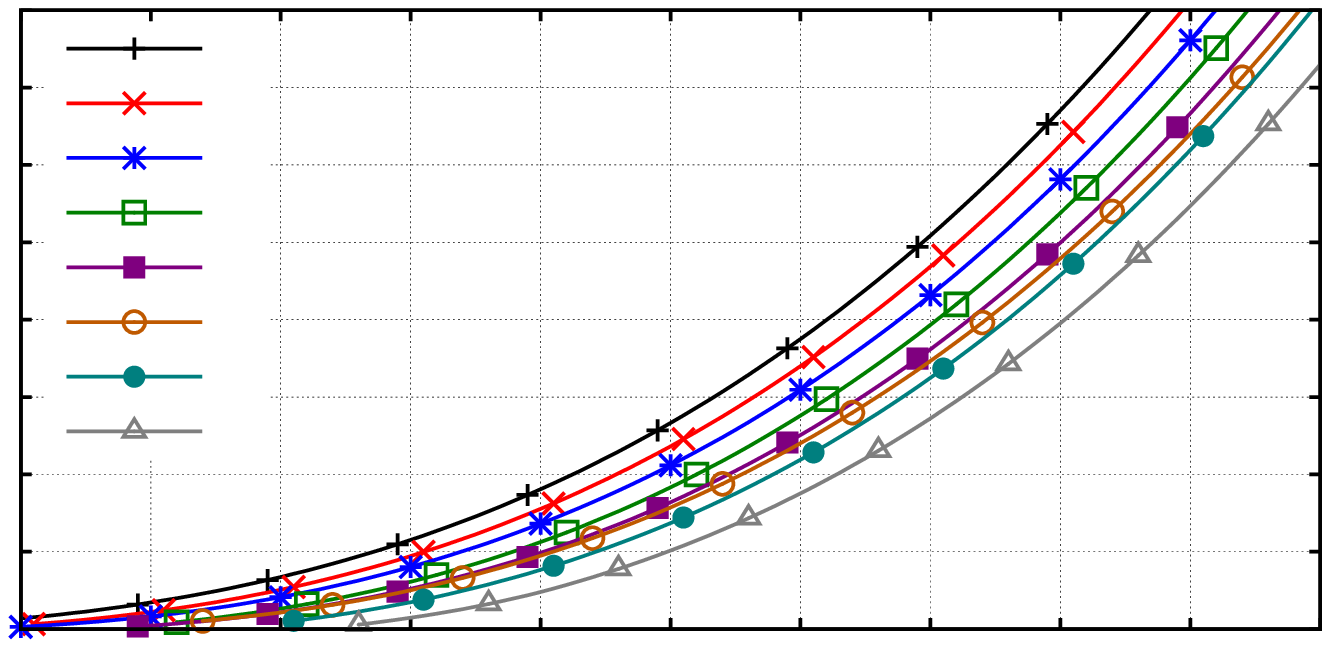}}%
    \gplfronttext
  \end{picture}%
\endgroup

%% file: err_0_0_0_0090.tex
% GNUPLOT: LaTeX picture with Postscript
\begingroup
  \makeatletter
  \providecommand\color[2][]{%
    \GenericError{(gnuplot) \space\space\space\@spaces}{%
      Package color not loaded in conjunction with
      terminal option `colourtext'%
    }{See the gnuplot documentation for explanation.%
    }{Either use 'blacktext' in gnuplot or load the package
      color.sty in LaTeX.}%
    \renewcommand\color[2][]{}%
  }%
  \providecommand\includegraphics[2][]{%
    \GenericError{(gnuplot) \space\space\space\@spaces}{%
      Package graphicx or graphics not loaded%
    }{See the gnuplot documentation for explanation.%
    }{The gnuplot epslatex terminal needs graphicx.sty or graphics.sty.}%
    \renewcommand\includegraphics[2][]{}%
  }%
  \providecommand\rotatebox[2]{#2}%
  \@ifundefined{ifGPcolor}{%
    \newif\ifGPcolor
    \GPcolortrue
  }{}%
  \@ifundefined{ifGPblacktext}{%
    \newif\ifGPblacktext
    \GPblacktexttrue
  }{}%
  % define a \g@addto@macro without @ in the name:
  \let\gplgaddtomacro\g@addto@macro
  % define empty templates for all commands taking text:
  \gdef\gplbacktext{}%
  \gdef\gplfronttext{}%
  \makeatother
  \ifGPblacktext
    % no textcolor at all
    \def\colorrgb#1{}%
    \def\colorgray#1{}%
  \else
    % gray or color?
    \ifGPcolor
      \def\colorrgb#1{\color[rgb]{#1}}%
      \def\colorgray#1{\color[gray]{#1}}%
      \expandafter\def\csname LTw\endcsname{\color{white}}%
      \expandafter\def\csname LTb\endcsname{\color{black}}%
      \expandafter\def\csname LTa\endcsname{\color{black}}%
      \expandafter\def\csname LT0\endcsname{\color[rgb]{1,0,0}}%
      \expandafter\def\csname LT1\endcsname{\color[rgb]{0,1,0}}%
      \expandafter\def\csname LT2\endcsname{\color[rgb]{0,0,1}}%
      \expandafter\def\csname LT3\endcsname{\color[rgb]{1,0,1}}%
      \expandafter\def\csname LT4\endcsname{\color[rgb]{0,1,1}}%
      \expandafter\def\csname LT5\endcsname{\color[rgb]{1,1,0}}%
      \expandafter\def\csname LT6\endcsname{\color[rgb]{0,0,0}}%
      \expandafter\def\csname LT7\endcsname{\color[rgb]{1,0.3,0}}%
      \expandafter\def\csname LT8\endcsname{\color[rgb]{0.5,0.5,0.5}}%
    \else
      % gray
      \def\colorrgb#1{\color{black}}%
      \def\colorgray#1{\color[gray]{#1}}%
      \expandafter\def\csname LTw\endcsname{\color{white}}%
      \expandafter\def\csname LTb\endcsname{\color{black}}%
      \expandafter\def\csname LTa\endcsname{\color{black}}%
      \expandafter\def\csname LT0\endcsname{\color{black}}%
      \expandafter\def\csname LT1\endcsname{\color{black}}%
      \expandafter\def\csname LT2\endcsname{\color{black}}%
      \expandafter\def\csname LT3\endcsname{\color{black}}%
      \expandafter\def\csname LT4\endcsname{\color{black}}%
      \expandafter\def\csname LT5\endcsname{\color{black}}%
      \expandafter\def\csname LT6\endcsname{\color{black}}%
      \expandafter\def\csname LT7\endcsname{\color{black}}%
      \expandafter\def\csname LT8\endcsname{\color{black}}%
    \fi
  \fi
    \setlength{\unitlength}{0.0500bp}%
    \ifx\gptboxheight\undefined%
      \newlength{\gptboxheight}%
      \newlength{\gptboxwidth}%
      \newsavebox{\gptboxtext}%
    \fi%
    \setlength{\fboxrule}{0.5pt}%
    \setlength{\fboxsep}{1pt}%
\begin{picture}(4534.00,4534.00)%
    \gplgaddtomacro\gplbacktext{%
      \csname LTb\endcsname%
      \put(607,704){\makebox(0,0)[r]{\strut{}-10}}%
      \put(607,1298){\makebox(0,0)[r]{\strut{}-8}}%
      \put(607,1892){\makebox(0,0)[r]{\strut{}-6}}%
      \put(607,2487){\makebox(0,0)[r]{\strut{}-4}}%
      \put(607,3081){\makebox(0,0)[r]{\strut{}-2}}%
      \put(607,3675){\makebox(0,0)[r]{\strut{}0}}%
      \put(607,4269){\makebox(0,0)[r]{\strut{}2}}%
      \put(739,484){\makebox(0,0){\strut{}$0.001$}}%
      \put(1216,484){\makebox(0,0){\strut{}$0.01$}}%
      \put(1693,484){\makebox(0,0){\strut{}$0.1$}}%
      \put(2171,484){\makebox(0,0){\strut{}$1$}}%
      \put(2648,484){\makebox(0,0){\strut{}$10$}}%
      \put(3125,484){\makebox(0,0){\strut{}$100$}}%
      \put(3257,704){\makebox(0,0)[l]{\strut{}$0.99$}}%
      \put(3257,1595){\makebox(0,0)[l]{\strut{}$0.995$}}%
      \put(3257,2487){\makebox(0,0)[l]{\strut{}$1$}}%
      \put(3257,3378){\makebox(0,0)[l]{\strut{}$1.005$}}%
      \put(3257,4269){\makebox(0,0)[l]{\strut{}$1.01$}}%
      \put(0,4488){\makebox(0,0)[l]{\strut{}$10^{x}$}}%
    }%
    \gplgaddtomacro\gplfronttext{%
      \csname LTb\endcsname%
      \put(176,2486){\rotatebox{-270}{\makebox(0,0){\strut{}$\kappa$}}}%
      \put(4026,2486){\rotatebox{-270}{\makebox(0,0){\strut{}vis}}}%
      \put(1932,154){\makebox(0,0){\strut{}$t$}}%
      \csname LTb\endcsname%
      \put(2138,1317){\makebox(0,0)[r]{\strut{}A0}}%
      \csname LTb\endcsname%
      \put(2138,1097){\makebox(0,0)[r]{\strut{}vis A0}}%
      \csname LTb\endcsname%
      \put(2138,877){\makebox(0,0)[r]{\strut{}power fit}}%
    }%
    \gplbacktext
    \put(0,0){\includegraphics{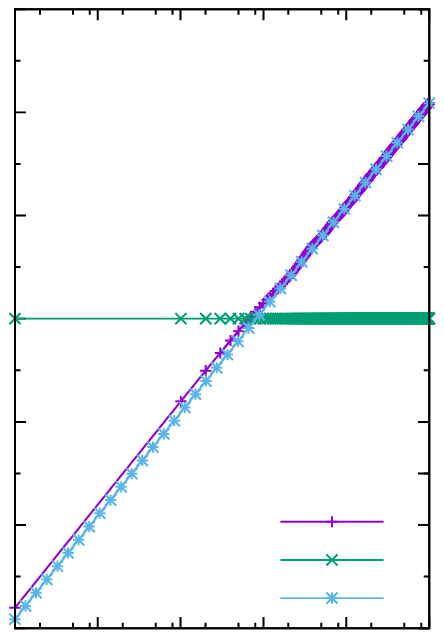}}%
    \gplfronttext
  \end{picture}%
\endgroup

%% file: err_1_0_0_0090.tex
% GNUPLOT: LaTeX picture with Postscript
\begingroup
  \makeatletter
  \providecommand\color[2][]{%
    \GenericError{(gnuplot) \space\space\space\@spaces}{%
      Package color not loaded in conjunction with
      terminal option `colourtext'%
    }{See the gnuplot documentation for explanation.%
    }{Either use 'blacktext' in gnuplot or load the package
      color.sty in LaTeX.}%
    \renewcommand\color[2][]{}%
  }%
  \providecommand\includegraphics[2][]{%
    \GenericError{(gnuplot) \space\space\space\@spaces}{%
      Package graphicx or graphics not loaded%
    }{See the gnuplot documentation for explanation.%
    }{The gnuplot epslatex terminal needs graphicx.sty or graphics.sty.}%
    \renewcommand\includegraphics[2][]{}%
  }%
  \providecommand\rotatebox[2]{#2}%
  \@ifundefined{ifGPcolor}{%
    \newif\ifGPcolor
    \GPcolortrue
  }{}%
  \@ifundefined{ifGPblacktext}{%
    \newif\ifGPblacktext
    \GPblacktexttrue
  }{}%
  % define a \g@addto@macro without @ in the name:
  \let\gplgaddtomacro\g@addto@macro
  % define empty templates for all commands taking text:
  \gdef\gplbacktext{}%
  \gdef\gplfronttext{}%
  \makeatother
  \ifGPblacktext
    % no textcolor at all
    \def\colorrgb#1{}%
    \def\colorgray#1{}%
  \else
    % gray or color?
    \ifGPcolor
      \def\colorrgb#1{\color[rgb]{#1}}%
      \def\colorgray#1{\color[gray]{#1}}%
      \expandafter\def\csname LTw\endcsname{\color{white}}%
      \expandafter\def\csname LTb\endcsname{\color{black}}%
      \expandafter\def\csname LTa\endcsname{\color{black}}%
      \expandafter\def\csname LT0\endcsname{\color[rgb]{1,0,0}}%
      \expandafter\def\csname LT1\endcsname{\color[rgb]{0,1,0}}%
      \expandafter\def\csname LT2\endcsname{\color[rgb]{0,0,1}}%
      \expandafter\def\csname LT3\endcsname{\color[rgb]{1,0,1}}%
      \expandafter\def\csname LT4\endcsname{\color[rgb]{0,1,1}}%
      \expandafter\def\csname LT5\endcsname{\color[rgb]{1,1,0}}%
      \expandafter\def\csname LT6\endcsname{\color[rgb]{0,0,0}}%
      \expandafter\def\csname LT7\endcsname{\color[rgb]{1,0.3,0}}%
      \expandafter\def\csname LT8\endcsname{\color[rgb]{0.5,0.5,0.5}}%
    \else
      % gray
      \def\colorrgb#1{\color{black}}%
      \def\colorgray#1{\color[gray]{#1}}%
      \expandafter\def\csname LTw\endcsname{\color{white}}%
      \expandafter\def\csname LTb\endcsname{\color{black}}%
      \expandafter\def\csname LTa\endcsname{\color{black}}%
      \expandafter\def\csname LT0\endcsname{\color{black}}%
      \expandafter\def\csname LT1\endcsname{\color{black}}%
      \expandafter\def\csname LT2\endcsname{\color{black}}%
      \expandafter\def\csname LT3\endcsname{\color{black}}%
      \expandafter\def\csname LT4\endcsname{\color{black}}%
      \expandafter\def\csname LT5\endcsname{\color{black}}%
      \expandafter\def\csname LT6\endcsname{\color{black}}%
      \expandafter\def\csname LT7\endcsname{\color{black}}%
      \expandafter\def\csname LT8\endcsname{\color{black}}%
    \fi
  \fi
    \setlength{\unitlength}{0.0500bp}%
    \ifx\gptboxheight\undefined%
      \newlength{\gptboxheight}%
      \newlength{\gptboxwidth}%
      \newsavebox{\gptboxtext}%
    \fi%
    \setlength{\fboxrule}{0.5pt}%
    \setlength{\fboxsep}{1pt}%
\begin{picture}(4534.00,4534.00)%
    \gplgaddtomacro\gplbacktext{%
      \csname LTb\endcsname%
      \put(814,704){\makebox(0,0)[r]{\strut{}-10}}%
      \put(814,1213){\makebox(0,0)[r]{\strut{}-8}}%
      \put(814,1723){\makebox(0,0)[r]{\strut{}-6}}%
      \put(814,2232){\makebox(0,0)[r]{\strut{}-4}}%
      \put(814,2741){\makebox(0,0)[r]{\strut{}-2}}%
      \put(814,3250){\makebox(0,0)[r]{\strut{}0}}%
      \put(814,3760){\makebox(0,0)[r]{\strut{}2}}%
      \put(814,4269){\makebox(0,0)[r]{\strut{}4}}%
      \put(946,484){\makebox(0,0){\strut{}$0$}}%
      \put(1461,484){\makebox(0,0){\strut{}$2$}}%
      \put(1976,484){\makebox(0,0){\strut{}$4$}}%
      \put(2491,484){\makebox(0,0){\strut{}$6$}}%
      \put(3006,484){\makebox(0,0){\strut{}$8$}}%
      \put(3521,484){\makebox(0,0){\strut{}$10$}}%
      \put(3653,704){\makebox(0,0)[l]{\strut{}-6}}%
      \put(3653,1298){\makebox(0,0)[l]{\strut{}-5}}%
      \put(3653,1892){\makebox(0,0)[l]{\strut{}-4}}%
      \put(3653,2487){\makebox(0,0)[l]{\strut{}-3}}%
      \put(3653,3081){\makebox(0,0)[l]{\strut{}-2}}%
      \put(3653,3675){\makebox(0,0)[l]{\strut{}-1}}%
      \put(3653,4269){\makebox(0,0)[l]{\strut{}0}}%
      \put(0,4488){\makebox(0,0)[l]{\strut{}$10^{x}$}}%
      \put(4034,4488){\makebox(0,0)[l]{\strut{}$10^{x}$}}%
    }%
    \gplgaddtomacro\gplfronttext{%
      \csname LTb\endcsname%
      \put(176,2486){\rotatebox{-270}{\makebox(0,0){\strut{}$\kappa$}}}%
      \put(4026,2486){\rotatebox{-270}{\makebox(0,0){\strut{}$\vert 1 - \text{vis} \vert$}}}%
      \put(2233,154){\makebox(0,0){\strut{}$t$}}%
      \csname LTb\endcsname%
      \put(2534,1537){\makebox(0,0)[r]{\strut{}A1}}%
      \csname LTb\endcsname%
      \put(2534,1317){\makebox(0,0)[r]{\strut{}$\vert 1 - \text{vis} \vert$}}%
      \csname LTb\endcsname%
      \put(2534,1097){\makebox(0,0)[r]{\strut{}power fit}}%
      \csname LTb\endcsname%
      \put(2534,877){\makebox(0,0)[r]{\strut{}exp fit}}%
    }%
    \gplbacktext
    \put(0,0){\includegraphics{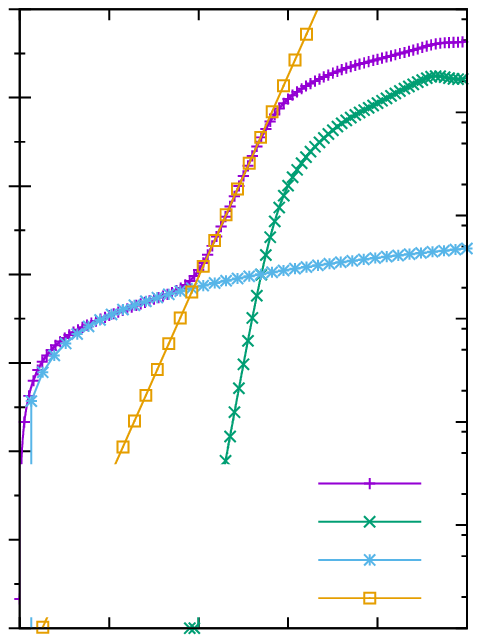}}%
    \gplfronttext
  \end{picture}%
\endgroup

%% file: decay_0.tex
% GNUPLOT: LaTeX picture with Postscript
\begingroup
  \makeatletter
  \providecommand\color[2][]{%
    \GenericError{(gnuplot) \space\space\space\@spaces}{%
      Package color not loaded in conjunction with
      terminal option `colourtext'%
    }{See the gnuplot documentation for explanation.%
    }{Either use 'blacktext' in gnuplot or load the package
      color.sty in LaTeX.}%
    \renewcommand\color[2][]{}%
  }%
  \providecommand\includegraphics[2][]{%
    \GenericError{(gnuplot) \space\space\space\@spaces}{%
      Package graphicx or graphics not loaded%
    }{See the gnuplot documentation for explanation.%
    }{The gnuplot epslatex terminal needs graphicx.sty or graphics.sty.}%
    \renewcommand\includegraphics[2][]{}%
  }%
  \providecommand\rotatebox[2]{#2}%
  \@ifundefined{ifGPcolor}{%
    \newif\ifGPcolor
    \GPcolortrue
  }{}%
  \@ifundefined{ifGPblacktext}{%
    \newif\ifGPblacktext
    \GPblacktexttrue
  }{}%
  % define a \g@addto@macro without @ in the name:
  \let\gplgaddtomacro\g@addto@macro
  % define empty templates for all commands taking text:
  \gdef\gplbacktext{}%
  \gdef\gplfronttext{}%
  \makeatother
  \ifGPblacktext
    % no textcolor at all
    \def\colorrgb#1{}%
    \def\colorgray#1{}%
  \else
    % gray or color?
    \ifGPcolor
      \def\colorrgb#1{\color[rgb]{#1}}%
      \def\colorgray#1{\color[gray]{#1}}%
      \expandafter\def\csname LTw\endcsname{\color{white}}%
      \expandafter\def\csname LTb\endcsname{\color{black}}%
      \expandafter\def\csname LTa\endcsname{\color{black}}%
      \expandafter\def\csname LT0\endcsname{\color[rgb]{1,0,0}}%
      \expandafter\def\csname LT1\endcsname{\color[rgb]{0,1,0}}%
      \expandafter\def\csname LT2\endcsname{\color[rgb]{0,0,1}}%
      \expandafter\def\csname LT3\endcsname{\color[rgb]{1,0,1}}%
      \expandafter\def\csname LT4\endcsname{\color[rgb]{0,1,1}}%
      \expandafter\def\csname LT5\endcsname{\color[rgb]{1,1,0}}%
      \expandafter\def\csname LT6\endcsname{\color[rgb]{0,0,0}}%
      \expandafter\def\csname LT7\endcsname{\color[rgb]{1,0.3,0}}%
      \expandafter\def\csname LT8\endcsname{\color[rgb]{0.5,0.5,0.5}}%
    \else
      % gray
      \def\colorrgb#1{\color{black}}%
      \def\colorgray#1{\color[gray]{#1}}%
      \expandafter\def\csname LTw\endcsname{\color{white}}%
      \expandafter\def\csname LTb\endcsname{\color{black}}%
      \expandafter\def\csname LTa\endcsname{\color{black}}%
      \expandafter\def\csname LT0\endcsname{\color{black}}%
      \expandafter\def\csname LT1\endcsname{\color{black}}%
      \expandafter\def\csname LT2\endcsname{\color{black}}%
      \expandafter\def\csname LT3\endcsname{\color{black}}%
      \expandafter\def\csname LT4\endcsname{\color{black}}%
      \expandafter\def\csname LT5\endcsname{\color{black}}%
      \expandafter\def\csname LT6\endcsname{\color{black}}%
      \expandafter\def\csname LT7\endcsname{\color{black}}%
      \expandafter\def\csname LT8\endcsname{\color{black}}%
    \fi
  \fi
    \setlength{\unitlength}{0.0500bp}%
    \ifx\gptboxheight\undefined%
      \newlength{\gptboxheight}%
      \newlength{\gptboxwidth}%
      \newsavebox{\gptboxtext}%
    \fi%
    \setlength{\fboxrule}{0.5pt}%
    \setlength{\fboxsep}{1pt}%
\begin{picture}(4534.00,4534.00)%
    \gplgaddtomacro\gplbacktext{%
      \csname LTb\endcsname%
      \put(616,990){\makebox(0,0)[r]{\strut{}$0.1$}}%
      \csname LTb\endcsname%
      \put(616,2630){\makebox(0,0)[r]{\strut{}$1$}}%
      \csname LTb\endcsname%
      \put(616,4269){\makebox(0,0)[r]{\strut{}$10$}}%
      \csname LTb\endcsname%
      \put(916,353){\rotatebox{70}{\makebox(0,0)[l]{\strut{}$2000$}}}%
      \csname LTb\endcsname%
      \put(1304,353){\rotatebox{70}{\makebox(0,0)[l]{\strut{}$4000$}}}%
      \csname LTb\endcsname%
      \put(1691,353){\rotatebox{70}{\makebox(0,0)[l]{\strut{}$6000$}}}%
      \csname LTb\endcsname%
      \put(2079,353){\rotatebox{70}{\makebox(0,0)[l]{\strut{}$8000$}}}%
      \csname LTb\endcsname%
      \put(2466,353){\rotatebox{70}{\makebox(0,0)[l]{\strut{}$10000$}}}%
      \csname LTb\endcsname%
      \put(2854,353){\rotatebox{70}{\makebox(0,0)[l]{\strut{}$12000$}}}%
      \csname LTb\endcsname%
      \put(3242,353){\rotatebox{70}{\makebox(0,0)[l]{\strut{}$14000$}}}%
      \csname LTb\endcsname%
      \put(3629,353){\rotatebox{70}{\makebox(0,0)[l]{\strut{}$16000$}}}%
    }%
    \gplgaddtomacro\gplfronttext{%
      \csname LTb\endcsname%
      \put(176,2629){\rotatebox{-270}{\makebox(0,0){\strut{}$\tau_{1/2}$}}}%
      \put(2442,154){\makebox(0,0){\strut{}$N$}}%
      \csname LTb\endcsname%
      \put(3150,4096){\makebox(0,0)[r]{\strut{}A3}}%
      \csname LTb\endcsname%
      \put(3150,3876){\makebox(0,0)[r]{\strut{}A1}}%
      \csname LTb\endcsname%
      \put(3150,3656){\makebox(0,0)[r]{\strut{}A2}}%
    }%
    \gplbacktext
    \put(0,0){\includegraphics{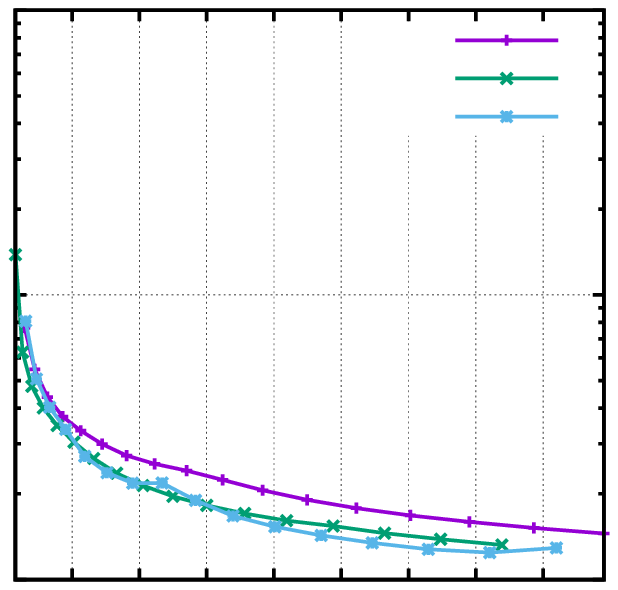}}%
    \gplfronttext
  \end{picture}%
\endgroup

%% file: decay_1.tex
% GNUPLOT: LaTeX picture with Postscript
\begingroup
  \makeatletter
  \providecommand\color[2][]{%
    \GenericError{(gnuplot) \space\space\space\@spaces}{%
      Package color not loaded in conjunction with
      terminal option `colourtext'%
    }{See the gnuplot documentation for explanation.%
    }{Either use 'blacktext' in gnuplot or load the package
      color.sty in LaTeX.}%
    \renewcommand\color[2][]{}%
  }%
  \providecommand\includegraphics[2][]{%
    \GenericError{(gnuplot) \space\space\space\@spaces}{%
      Package graphicx or graphics not loaded%
    }{See the gnuplot documentation for explanation.%
    }{The gnuplot epslatex terminal needs graphicx.sty or graphics.sty.}%
    \renewcommand\includegraphics[2][]{}%
  }%
  \providecommand\rotatebox[2]{#2}%
  \@ifundefined{ifGPcolor}{%
    \newif\ifGPcolor
    \GPcolortrue
  }{}%
  \@ifundefined{ifGPblacktext}{%
    \newif\ifGPblacktext
    \GPblacktexttrue
  }{}%
  % define a \g@addto@macro without @ in the name:
  \let\gplgaddtomacro\g@addto@macro
  % define empty templates for all commands taking text:
  \gdef\gplbacktext{}%
  \gdef\gplfronttext{}%
  \makeatother
  \ifGPblacktext
    % no textcolor at all
    \def\colorrgb#1{}%
    \def\colorgray#1{}%
  \else
    % gray or color?
    \ifGPcolor
      \def\colorrgb#1{\color[rgb]{#1}}%
      \def\colorgray#1{\color[gray]{#1}}%
      \expandafter\def\csname LTw\endcsname{\color{white}}%
      \expandafter\def\csname LTb\endcsname{\color{black}}%
      \expandafter\def\csname LTa\endcsname{\color{black}}%
      \expandafter\def\csname LT0\endcsname{\color[rgb]{1,0,0}}%
      \expandafter\def\csname LT1\endcsname{\color[rgb]{0,1,0}}%
      \expandafter\def\csname LT2\endcsname{\color[rgb]{0,0,1}}%
      \expandafter\def\csname LT3\endcsname{\color[rgb]{1,0,1}}%
      \expandafter\def\csname LT4\endcsname{\color[rgb]{0,1,1}}%
      \expandafter\def\csname LT5\endcsname{\color[rgb]{1,1,0}}%
      \expandafter\def\csname LT6\endcsname{\color[rgb]{0,0,0}}%
      \expandafter\def\csname LT7\endcsname{\color[rgb]{1,0.3,0}}%
      \expandafter\def\csname LT8\endcsname{\color[rgb]{0.5,0.5,0.5}}%
    \else
      % gray
      \def\colorrgb#1{\color{black}}%
      \def\colorgray#1{\color[gray]{#1}}%
      \expandafter\def\csname LTw\endcsname{\color{white}}%
      \expandafter\def\csname LTb\endcsname{\color{black}}%
      \expandafter\def\csname LTa\endcsname{\color{black}}%
      \expandafter\def\csname LT0\endcsname{\color{black}}%
      \expandafter\def\csname LT1\endcsname{\color{black}}%
      \expandafter\def\csname LT2\endcsname{\color{black}}%
      \expandafter\def\csname LT3\endcsname{\color{black}}%
      \expandafter\def\csname LT4\endcsname{\color{black}}%
      \expandafter\def\csname LT5\endcsname{\color{black}}%
      \expandafter\def\csname LT6\endcsname{\color{black}}%
      \expandafter\def\csname LT7\endcsname{\color{black}}%
      \expandafter\def\csname LT8\endcsname{\color{black}}%
    \fi
  \fi
    \setlength{\unitlength}{0.0500bp}%
    \ifx\gptboxheight\undefined%
      \newlength{\gptboxheight}%
      \newlength{\gptboxwidth}%
      \newsavebox{\gptboxtext}%
    \fi%
    \setlength{\fboxrule}{0.5pt}%
    \setlength{\fboxsep}{1pt}%
\begin{picture}(4534.00,4534.00)%
    \gplgaddtomacro\gplbacktext{%
      \csname LTb\endcsname%
      \put(616,990){\makebox(0,0)[r]{\strut{}$0.1$}}%
      \csname LTb\endcsname%
      \put(616,2630){\makebox(0,0)[r]{\strut{}$1$}}%
      \csname LTb\endcsname%
      \put(616,4269){\makebox(0,0)[r]{\strut{}$10$}}%
      \csname LTb\endcsname%
      \put(898,353){\rotatebox{70}{\makebox(0,0)[l]{\strut{}$2000$}}}%
      \csname LTb\endcsname%
      \put(1206,353){\rotatebox{70}{\makebox(0,0)[l]{\strut{}$4000$}}}%
      \csname LTb\endcsname%
      \put(1514,353){\rotatebox{70}{\makebox(0,0)[l]{\strut{}$6000$}}}%
      \csname LTb\endcsname%
      \put(1822,353){\rotatebox{70}{\makebox(0,0)[l]{\strut{}$8000$}}}%
      \csname LTb\endcsname%
      \put(2130,353){\rotatebox{70}{\makebox(0,0)[l]{\strut{}$10000$}}}%
      \csname LTb\endcsname%
      \put(2439,353){\rotatebox{70}{\makebox(0,0)[l]{\strut{}$12000$}}}%
      \csname LTb\endcsname%
      \put(2747,353){\rotatebox{70}{\makebox(0,0)[l]{\strut{}$14000$}}}%
      \csname LTb\endcsname%
      \put(3055,353){\rotatebox{70}{\makebox(0,0)[l]{\strut{}$16000$}}}%
      \csname LTb\endcsname%
      \put(3363,353){\rotatebox{70}{\makebox(0,0)[l]{\strut{}$18000$}}}%
      \csname LTb\endcsname%
      \put(3671,353){\rotatebox{70}{\makebox(0,0)[l]{\strut{}$20000$}}}%
    }%
    \gplgaddtomacro\gplfronttext{%
      \csname LTb\endcsname%
      \put(176,2629){\rotatebox{-270}{\makebox(0,0){\strut{}$\tau_{1/2}$}}}%
      \put(2442,154){\makebox(0,0){\strut{}$N$}}%
      \csname LTb\endcsname%
      \put(3150,4096){\makebox(0,0)[r]{\strut{}B3}}%
      \csname LTb\endcsname%
      \put(3150,3876){\makebox(0,0)[r]{\strut{}B1}}%
      \csname LTb\endcsname%
      \put(3150,3656){\makebox(0,0)[r]{\strut{}B2}}%
    }%
    \gplbacktext
    \put(0,0){\includegraphics{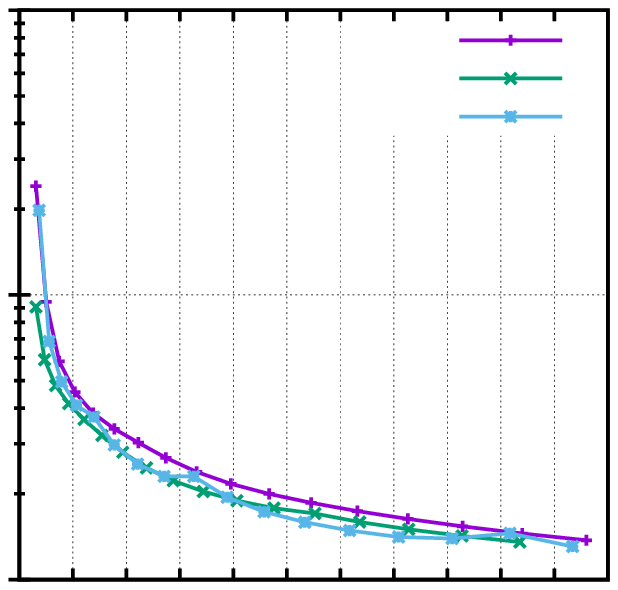}}%
    \gplfronttext
  \end{picture}%
\endgroup

%% file: cross_section_demo_010_0050.tex
% GNUPLOT: LaTeX picture with Postscript
\begingroup
  \makeatletter
  \providecommand\color[2][]{%
    \GenericError{(gnuplot) \space\space\space\@spaces}{%
      Package color not loaded in conjunction with
      terminal option `colourtext'%
    }{See the gnuplot documentation for explanation.%
    }{Either use 'blacktext' in gnuplot or load the package
      color.sty in LaTeX.}%
    \renewcommand\color[2][]{}%
  }%
  \providecommand\includegraphics[2][]{%
    \GenericError{(gnuplot) \space\space\space\@spaces}{%
      Package graphicx or graphics not loaded%
    }{See the gnuplot documentation for explanation.%
    }{The gnuplot epslatex terminal needs graphicx.sty or graphics.sty.}%
    \renewcommand\includegraphics[2][]{}%
  }%
  \providecommand\rotatebox[2]{#2}%
  \@ifundefined{ifGPcolor}{%
    \newif\ifGPcolor
    \GPcolortrue
  }{}%
  \@ifundefined{ifGPblacktext}{%
    \newif\ifGPblacktext
    \GPblacktexttrue
  }{}%
  % define a \g@addto@macro without @ in the name:
  \let\gplgaddtomacro\g@addto@macro
  % define empty templates for all commands taking text:
  \gdef\gplbacktext{}%
  \gdef\gplfronttext{}%
  \makeatother
  \ifGPblacktext
    % no textcolor at all
    \def\colorrgb#1{}%
    \def\colorgray#1{}%
  \else
    % gray or color?
    \ifGPcolor
      \def\colorrgb#1{\color[rgb]{#1}}%
      \def\colorgray#1{\color[gray]{#1}}%
      \expandafter\def\csname LTw\endcsname{\color{white}}%
      \expandafter\def\csname LTb\endcsname{\color{black}}%
      \expandafter\def\csname LTa\endcsname{\color{black}}%
      \expandafter\def\csname LT0\endcsname{\color[rgb]{1,0,0}}%
      \expandafter\def\csname LT1\endcsname{\color[rgb]{0,1,0}}%
      \expandafter\def\csname LT2\endcsname{\color[rgb]{0,0,1}}%
      \expandafter\def\csname LT3\endcsname{\color[rgb]{1,0,1}}%
      \expandafter\def\csname LT4\endcsname{\color[rgb]{0,1,1}}%
      \expandafter\def\csname LT5\endcsname{\color[rgb]{1,1,0}}%
      \expandafter\def\csname LT6\endcsname{\color[rgb]{0,0,0}}%
      \expandafter\def\csname LT7\endcsname{\color[rgb]{1,0.3,0}}%
      \expandafter\def\csname LT8\endcsname{\color[rgb]{0.5,0.5,0.5}}%
    \else
      % gray
      \def\colorrgb#1{\color{black}}%
      \def\colorgray#1{\color[gray]{#1}}%
      \expandafter\def\csname LTw\endcsname{\color{white}}%
      \expandafter\def\csname LTb\endcsname{\color{black}}%
      \expandafter\def\csname LTa\endcsname{\color{black}}%
      \expandafter\def\csname LT0\endcsname{\color{black}}%
      \expandafter\def\csname LT1\endcsname{\color{black}}%
      \expandafter\def\csname LT2\endcsname{\color{black}}%
      \expandafter\def\csname LT3\endcsname{\color{black}}%
      \expandafter\def\csname LT4\endcsname{\color{black}}%
      \expandafter\def\csname LT5\endcsname{\color{black}}%
      \expandafter\def\csname LT6\endcsname{\color{black}}%
      \expandafter\def\csname LT7\endcsname{\color{black}}%
      \expandafter\def\csname LT8\endcsname{\color{black}}%
    \fi
  \fi
    \setlength{\unitlength}{0.0500bp}%
    \ifx\gptboxheight\undefined%
      \newlength{\gptboxheight}%
      \newlength{\gptboxwidth}%
      \newsavebox{\gptboxtext}%
    \fi%
    \setlength{\fboxrule}{0.5pt}%
    \setlength{\fboxsep}{1pt}%
\begin{picture}(4534.00,3400.00)%
    \gplgaddtomacro\gplbacktext{%
      \csname LTb\endcsname%
      \put(682,704){\makebox(0,0)[r]{\strut{}$2$}}%
      \csname LTb\endcsname%
      \put(682,1051){\makebox(0,0)[r]{\strut{}$4$}}%
      \csname LTb\endcsname%
      \put(682,1399){\makebox(0,0)[r]{\strut{}$6$}}%
      \csname LTb\endcsname%
      \put(682,1746){\makebox(0,0)[r]{\strut{}$8$}}%
      \csname LTb\endcsname%
      \put(682,2093){\makebox(0,0)[r]{\strut{}$10$}}%
      \csname LTb\endcsname%
      \put(682,2440){\makebox(0,0)[r]{\strut{}$12$}}%
      \csname LTb\endcsname%
      \put(682,2788){\makebox(0,0)[r]{\strut{}$14$}}%
      \csname LTb\endcsname%
      \put(682,3135){\makebox(0,0)[r]{\strut{}$16$}}%
      \csname LTb\endcsname%
      \put(1418,484){\makebox(0,0){\strut{}$2.4$}}%
      \csname LTb\endcsname%
      \put(2098,484){\makebox(0,0){\strut{}$2.6$}}%
      \csname LTb\endcsname%
      \put(2778,484){\makebox(0,0){\strut{}$2.8$}}%
      \csname LTb\endcsname%
      \put(3457,484){\makebox(0,0){\strut{}$3$}}%
      \csname LTb\endcsname%
      \put(4137,484){\makebox(0,0){\strut{}$3.2$}}%
    }%
    \gplgaddtomacro\gplfronttext{%
      \csname LTb\endcsname%
      \put(176,1919){\rotatebox{-270}{\makebox(0,0){\strut{}${\rm spp}(z)$}}}%
      \put(2475,154){\makebox(0,0){\strut{}$x$}}%
    }%
    \gplbacktext
    \put(0,0){\includegraphics{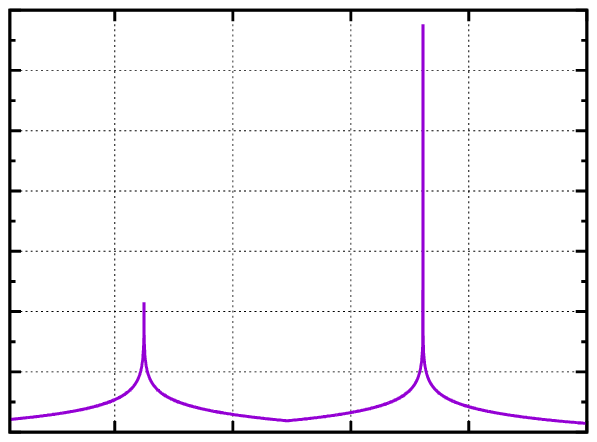}}%
    \gplfronttext
  \end{picture}%
\endgroup

%% file: decay_A1_cmp.tex
% GNUPLOT: LaTeX picture with Postscript
\begingroup
  \makeatletter
  \providecommand\color[2][]{%
    \GenericError{(gnuplot) \space\space\space\@spaces}{%
      Package color not loaded in conjunction with
      terminal option `colourtext'%
    }{See the gnuplot documentation for explanation.%
    }{Either use 'blacktext' in gnuplot or load the package
      color.sty in LaTeX.}%
    \renewcommand\color[2][]{}%
  }%
  \providecommand\includegraphics[2][]{%
    \GenericError{(gnuplot) \space\space\space\@spaces}{%
      Package graphicx or graphics not loaded%
    }{See the gnuplot documentation for explanation.%
    }{The gnuplot epslatex terminal needs graphicx.sty or graphics.sty.}%
    \renewcommand\includegraphics[2][]{}%
  }%
  \providecommand\rotatebox[2]{#2}%
  \@ifundefined{ifGPcolor}{%
    \newif\ifGPcolor
    \GPcolortrue
  }{}%
  \@ifundefined{ifGPblacktext}{%
    \newif\ifGPblacktext
    \GPblacktexttrue
  }{}%
  % define a \g@addto@macro without @ in the name:
  \let\gplgaddtomacro\g@addto@macro
  % define empty templates for all commands taking text:
  \gdef\gplbacktext{}%
  \gdef\gplfronttext{}%
  \makeatother
  \ifGPblacktext
    % no textcolor at all
    \def\colorrgb#1{}%
    \def\colorgray#1{}%
  \else
    % gray or color?
    \ifGPcolor
      \def\colorrgb#1{\color[rgb]{#1}}%
      \def\colorgray#1{\color[gray]{#1}}%
      \expandafter\def\csname LTw\endcsname{\color{white}}%
      \expandafter\def\csname LTb\endcsname{\color{black}}%
      \expandafter\def\csname LTa\endcsname{\color{black}}%
      \expandafter\def\csname LT0\endcsname{\color[rgb]{1,0,0}}%
      \expandafter\def\csname LT1\endcsname{\color[rgb]{0,1,0}}%
      \expandafter\def\csname LT2\endcsname{\color[rgb]{0,0,1}}%
      \expandafter\def\csname LT3\endcsname{\color[rgb]{1,0,1}}%
      \expandafter\def\csname LT4\endcsname{\color[rgb]{0,1,1}}%
      \expandafter\def\csname LT5\endcsname{\color[rgb]{1,1,0}}%
      \expandafter\def\csname LT6\endcsname{\color[rgb]{0,0,0}}%
      \expandafter\def\csname LT7\endcsname{\color[rgb]{1,0.3,0}}%
      \expandafter\def\csname LT8\endcsname{\color[rgb]{0.5,0.5,0.5}}%
    \else
      % gray
      \def\colorrgb#1{\color{black}}%
      \def\colorgray#1{\color[gray]{#1}}%
      \expandafter\def\csname LTw\endcsname{\color{white}}%
      \expandafter\def\csname LTb\endcsname{\color{black}}%
      \expandafter\def\csname LTa\endcsname{\color{black}}%
      \expandafter\def\csname LT0\endcsname{\color{black}}%
      \expandafter\def\csname LT1\endcsname{\color{black}}%
      \expandafter\def\csname LT2\endcsname{\color{black}}%
      \expandafter\def\csname LT3\endcsname{\color{black}}%
      \expandafter\def\csname LT4\endcsname{\color{black}}%
      \expandafter\def\csname LT5\endcsname{\color{black}}%
      \expandafter\def\csname LT6\endcsname{\color{black}}%
      \expandafter\def\csname LT7\endcsname{\color{black}}%
      \expandafter\def\csname LT8\endcsname{\color{black}}%
    \fi
  \fi
    \setlength{\unitlength}{0.0500bp}%
    \ifx\gptboxheight\undefined%
      \newlength{\gptboxheight}%
      \newlength{\gptboxwidth}%
      \newsavebox{\gptboxtext}%
    \fi%
    \setlength{\fboxrule}{0.5pt}%
    \setlength{\fboxsep}{1pt}%
\begin{picture}(4534.00,4534.00)%
    \gplgaddtomacro\gplbacktext{%
      \csname LTb\endcsname%
      \put(748,990){\makebox(0,0)[r]{\strut{}$0.01$}}%
      \csname LTb\endcsname%
      \put(748,1951){\makebox(0,0)[r]{\strut{}$0.1$}}%
      \csname LTb\endcsname%
      \put(748,2912){\makebox(0,0)[r]{\strut{}$1$}}%
      \csname LTb\endcsname%
      \put(748,3873){\makebox(0,0)[r]{\strut{}$10$}}%
      \csname LTb\endcsname%
      \put(1250,353){\rotatebox{70}{\makebox(0,0)[l]{\strut{}$2000$}}}%
      \csname LTb\endcsname%
      \put(1813,353){\rotatebox{70}{\makebox(0,0)[l]{\strut{}$4000$}}}%
      \csname LTb\endcsname%
      \put(2377,353){\rotatebox{70}{\makebox(0,0)[l]{\strut{}$6000$}}}%
      \csname LTb\endcsname%
      \put(2940,353){\rotatebox{70}{\makebox(0,0)[l]{\strut{}$8000$}}}%
      \csname LTb\endcsname%
      \put(3503,353){\rotatebox{70}{\makebox(0,0)[l]{\strut{}$10000$}}}%
    }%
    \gplgaddtomacro\gplfronttext{%
      \csname LTb\endcsname%
      \put(176,2431){\rotatebox{-270}{\makebox(0,0){\strut{}$\tau$}}}%
      \put(2508,154){\makebox(0,0){\strut{}$N$}}%
      \put(2508,4203){\makebox(0,0){\strut{}A1}}%
      \csname LTb\endcsname%
      \put(3150,3700){\makebox(0,0)[r]{\strut{}operator expansion}}%
      \csname LTb\endcsname%
      \put(3150,3480){\makebox(0,0)[r]{\strut{}real time}}%
    }%
    \gplbacktext
    \put(0,0){\includegraphics{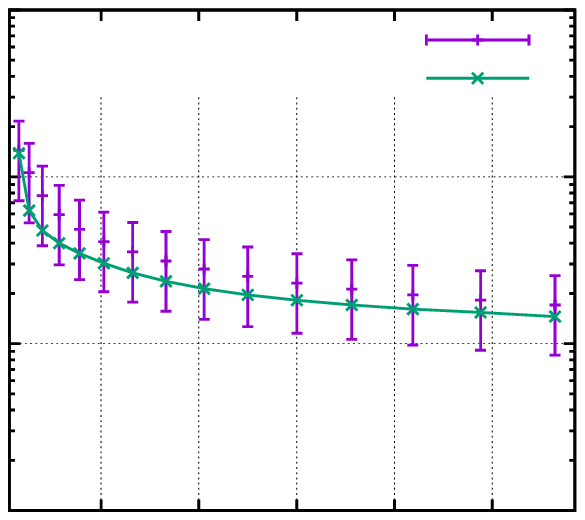}}%
    \gplfronttext
  \end{picture}%
\endgroup

%% file: decay_B1_cmp.tex
% GNUPLOT: LaTeX picture with Postscript
\begingroup
  \makeatletter
  \providecommand\color[2][]{%
    \GenericError{(gnuplot) \space\space\space\@spaces}{%
      Package color not loaded in conjunction with
      terminal option `colourtext'%
    }{See the gnuplot documentation for explanation.%
    }{Either use 'blacktext' in gnuplot or load the package
      color.sty in LaTeX.}%
    \renewcommand\color[2][]{}%
  }%
  \providecommand\includegraphics[2][]{%
    \GenericError{(gnuplot) \space\space\space\@spaces}{%
      Package graphicx or graphics not loaded%
    }{See the gnuplot documentation for explanation.%
    }{The gnuplot epslatex terminal needs graphicx.sty or graphics.sty.}%
    \renewcommand\includegraphics[2][]{}%
  }%
  \providecommand\rotatebox[2]{#2}%
  \@ifundefined{ifGPcolor}{%
    \newif\ifGPcolor
    \GPcolortrue
  }{}%
  \@ifundefined{ifGPblacktext}{%
    \newif\ifGPblacktext
    \GPblacktexttrue
  }{}%
  % define a \g@addto@macro without @ in the name:
  \let\gplgaddtomacro\g@addto@macro
  % define empty templates for all commands taking text:
  \gdef\gplbacktext{}%
  \gdef\gplfronttext{}%
  \makeatother
  \ifGPblacktext
    % no textcolor at all
    \def\colorrgb#1{}%
    \def\colorgray#1{}%
  \else
    % gray or color?
    \ifGPcolor
      \def\colorrgb#1{\color[rgb]{#1}}%
      \def\colorgray#1{\color[gray]{#1}}%
      \expandafter\def\csname LTw\endcsname{\color{white}}%
      \expandafter\def\csname LTb\endcsname{\color{black}}%
      \expandafter\def\csname LTa\endcsname{\color{black}}%
      \expandafter\def\csname LT0\endcsname{\color[rgb]{1,0,0}}%
      \expandafter\def\csname LT1\endcsname{\color[rgb]{0,1,0}}%
      \expandafter\def\csname LT2\endcsname{\color[rgb]{0,0,1}}%
      \expandafter\def\csname LT3\endcsname{\color[rgb]{1,0,1}}%
      \expandafter\def\csname LT4\endcsname{\color[rgb]{0,1,1}}%
      \expandafter\def\csname LT5\endcsname{\color[rgb]{1,1,0}}%
      \expandafter\def\csname LT6\endcsname{\color[rgb]{0,0,0}}%
      \expandafter\def\csname LT7\endcsname{\color[rgb]{1,0.3,0}}%
      \expandafter\def\csname LT8\endcsname{\color[rgb]{0.5,0.5,0.5}}%
    \else
      % gray
      \def\colorrgb#1{\color{black}}%
      \def\colorgray#1{\color[gray]{#1}}%
      \expandafter\def\csname LTw\endcsname{\color{white}}%
      \expandafter\def\csname LTb\endcsname{\color{black}}%
      \expandafter\def\csname LTa\endcsname{\color{black}}%
      \expandafter\def\csname LT0\endcsname{\color{black}}%
      \expandafter\def\csname LT1\endcsname{\color{black}}%
      \expandafter\def\csname LT2\endcsname{\color{black}}%
      \expandafter\def\csname LT3\endcsname{\color{black}}%
      \expandafter\def\csname LT4\endcsname{\color{black}}%
      \expandafter\def\csname LT5\endcsname{\color{black}}%
      \expandafter\def\csname LT6\endcsname{\color{black}}%
      \expandafter\def\csname LT7\endcsname{\color{black}}%
      \expandafter\def\csname LT8\endcsname{\color{black}}%
    \fi
  \fi
    \setlength{\unitlength}{0.0500bp}%
    \ifx\gptboxheight\undefined%
      \newlength{\gptboxheight}%
      \newlength{\gptboxwidth}%
      \newsavebox{\gptboxtext}%
    \fi%
    \setlength{\fboxrule}{0.5pt}%
    \setlength{\fboxsep}{1pt}%
\begin{picture}(4534.00,4534.00)%
    \gplgaddtomacro\gplbacktext{%
      \csname LTb\endcsname%
      \put(748,990){\makebox(0,0)[r]{\strut{}$0.01$}}%
      \csname LTb\endcsname%
      \put(748,1951){\makebox(0,0)[r]{\strut{}$0.1$}}%
      \csname LTb\endcsname%
      \put(748,2912){\makebox(0,0)[r]{\strut{}$1$}}%
      \csname LTb\endcsname%
      \put(748,3873){\makebox(0,0)[r]{\strut{}$10$}}%
      \csname LTb\endcsname%
      \put(1132,353){\rotatebox{70}{\makebox(0,0)[l]{\strut{}$2000$}}}%
      \csname LTb\endcsname%
      \put(1568,353){\rotatebox{70}{\makebox(0,0)[l]{\strut{}$4000$}}}%
      \csname LTb\endcsname%
      \put(2004,353){\rotatebox{70}{\makebox(0,0)[l]{\strut{}$6000$}}}%
      \csname LTb\endcsname%
      \put(2440,353){\rotatebox{70}{\makebox(0,0)[l]{\strut{}$8000$}}}%
      \csname LTb\endcsname%
      \put(2877,353){\rotatebox{70}{\makebox(0,0)[l]{\strut{}$10000$}}}%
      \csname LTb\endcsname%
      \put(3313,353){\rotatebox{70}{\makebox(0,0)[l]{\strut{}$12000$}}}%
      \csname LTb\endcsname%
      \put(3749,353){\rotatebox{70}{\makebox(0,0)[l]{\strut{}$14000$}}}%
    }%
    \gplgaddtomacro\gplfronttext{%
      \csname LTb\endcsname%
      \put(176,2431){\rotatebox{-270}{\makebox(0,0){\strut{}$\tau$}}}%
      \put(2508,154){\makebox(0,0){\strut{}$N$}}%
      \put(2508,4203){\makebox(0,0){\strut{}B1}}%
      \csname LTb\endcsname%
      \put(3150,3700){\makebox(0,0)[r]{\strut{}operator expansion}}%
      \csname LTb\endcsname%
      \put(3150,3480){\makebox(0,0)[r]{\strut{}real time}}%
    }%
    \gplbacktext
    \put(0,0){\includegraphics{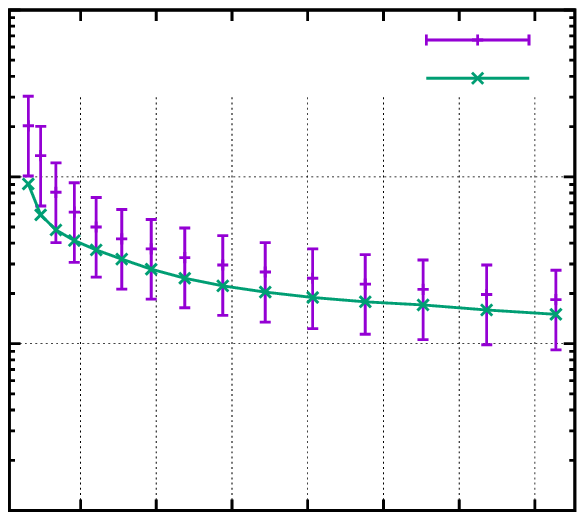}}%
    \gplfronttext
  \end{picture}%
\endgroup

%% file: decay_A3_cmp.tex
% GNUPLOT: LaTeX picture with Postscript
\begingroup
  \makeatletter
  \providecommand\color[2][]{%
    \GenericError{(gnuplot) \space\space\space\@spaces}{%
      Package color not loaded in conjunction with
      terminal option `colourtext'%
    }{See the gnuplot documentation for explanation.%
    }{Either use 'blacktext' in gnuplot or load the package
      color.sty in LaTeX.}%
    \renewcommand\color[2][]{}%
  }%
  \providecommand\includegraphics[2][]{%
    \GenericError{(gnuplot) \space\space\space\@spaces}{%
      Package graphicx or graphics not loaded%
    }{See the gnuplot documentation for explanation.%
    }{The gnuplot epslatex terminal needs graphicx.sty or graphics.sty.}%
    \renewcommand\includegraphics[2][]{}%
  }%
  \providecommand\rotatebox[2]{#2}%
  \@ifundefined{ifGPcolor}{%
    \newif\ifGPcolor
    \GPcolortrue
  }{}%
  \@ifundefined{ifGPblacktext}{%
    \newif\ifGPblacktext
    \GPblacktexttrue
  }{}%
  % define a \g@addto@macro without @ in the name:
  \let\gplgaddtomacro\g@addto@macro
  % define empty templates for all commands taking text:
  \gdef\gplbacktext{}%
  \gdef\gplfronttext{}%
  \makeatother
  \ifGPblacktext
    % no textcolor at all
    \def\colorrgb#1{}%
    \def\colorgray#1{}%
  \else
    % gray or color?
    \ifGPcolor
      \def\colorrgb#1{\color[rgb]{#1}}%
      \def\colorgray#1{\color[gray]{#1}}%
      \expandafter\def\csname LTw\endcsname{\color{white}}%
      \expandafter\def\csname LTb\endcsname{\color{black}}%
      \expandafter\def\csname LTa\endcsname{\color{black}}%
      \expandafter\def\csname LT0\endcsname{\color[rgb]{1,0,0}}%
      \expandafter\def\csname LT1\endcsname{\color[rgb]{0,1,0}}%
      \expandafter\def\csname LT2\endcsname{\color[rgb]{0,0,1}}%
      \expandafter\def\csname LT3\endcsname{\color[rgb]{1,0,1}}%
      \expandafter\def\csname LT4\endcsname{\color[rgb]{0,1,1}}%
      \expandafter\def\csname LT5\endcsname{\color[rgb]{1,1,0}}%
      \expandafter\def\csname LT6\endcsname{\color[rgb]{0,0,0}}%
      \expandafter\def\csname LT7\endcsname{\color[rgb]{1,0.3,0}}%
      \expandafter\def\csname LT8\endcsname{\color[rgb]{0.5,0.5,0.5}}%
    \else
      % gray
      \def\colorrgb#1{\color{black}}%
      \def\colorgray#1{\color[gray]{#1}}%
      \expandafter\def\csname LTw\endcsname{\color{white}}%
      \expandafter\def\csname LTb\endcsname{\color{black}}%
      \expandafter\def\csname LTa\endcsname{\color{black}}%
      \expandafter\def\csname LT0\endcsname{\color{black}}%
      \expandafter\def\csname LT1\endcsname{\color{black}}%
      \expandafter\def\csname LT2\endcsname{\color{black}}%
      \expandafter\def\csname LT3\endcsname{\color{black}}%
      \expandafter\def\csname LT4\endcsname{\color{black}}%
      \expandafter\def\csname LT5\endcsname{\color{black}}%
      \expandafter\def\csname LT6\endcsname{\color{black}}%
      \expandafter\def\csname LT7\endcsname{\color{black}}%
      \expandafter\def\csname LT8\endcsname{\color{black}}%
    \fi
  \fi
    \setlength{\unitlength}{0.0500bp}%
    \ifx\gptboxheight\undefined%
      \newlength{\gptboxheight}%
      \newlength{\gptboxwidth}%
      \newsavebox{\gptboxtext}%
    \fi%
    \setlength{\fboxrule}{0.5pt}%
    \setlength{\fboxsep}{1pt}%
\begin{picture}(4534.00,4534.00)%
    \gplgaddtomacro\gplbacktext{%
      \csname LTb\endcsname%
      \put(748,990){\makebox(0,0)[r]{\strut{}$0.01$}}%
      \csname LTb\endcsname%
      \put(748,1951){\makebox(0,0)[r]{\strut{}$0.1$}}%
      \csname LTb\endcsname%
      \put(748,2912){\makebox(0,0)[r]{\strut{}$1$}}%
      \csname LTb\endcsname%
      \put(748,3873){\makebox(0,0)[r]{\strut{}$10$}}%
      \csname LTb\endcsname%
      \put(1159,353){\rotatebox{70}{\makebox(0,0)[l]{\strut{}$2000$}}}%
      \csname LTb\endcsname%
      \put(1617,353){\rotatebox{70}{\makebox(0,0)[l]{\strut{}$4000$}}}%
      \csname LTb\endcsname%
      \put(2075,353){\rotatebox{70}{\makebox(0,0)[l]{\strut{}$6000$}}}%
      \csname LTb\endcsname%
      \put(2533,353){\rotatebox{70}{\makebox(0,0)[l]{\strut{}$8000$}}}%
      \csname LTb\endcsname%
      \put(2991,353){\rotatebox{70}{\makebox(0,0)[l]{\strut{}$10000$}}}%
      \csname LTb\endcsname%
      \put(3449,353){\rotatebox{70}{\makebox(0,0)[l]{\strut{}$12000$}}}%
      \csname LTb\endcsname%
      \put(3908,353){\rotatebox{70}{\makebox(0,0)[l]{\strut{}$14000$}}}%
    }%
    \gplgaddtomacro\gplfronttext{%
      \csname LTb\endcsname%
      \put(176,2431){\rotatebox{-270}{\makebox(0,0){\strut{}$\tau$}}}%
      \put(2508,154){\makebox(0,0){\strut{}$N$}}%
      \put(2508,4203){\makebox(0,0){\strut{}A3}}%
      \csname LTb\endcsname%
      \put(3150,3700){\makebox(0,0)[r]{\strut{}operator expansion}}%
      \csname LTb\endcsname%
      \put(3150,3480){\makebox(0,0)[r]{\strut{}real time}}%
    }%
    \gplbacktext
    \put(0,0){\includegraphics{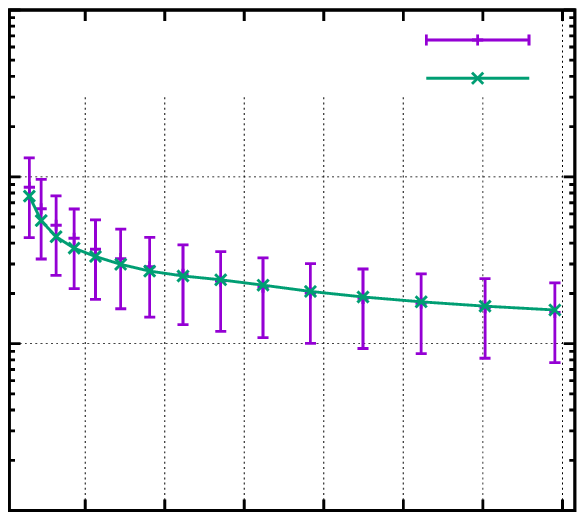}}%
    \gplfronttext
  \end{picture}%
\endgroup

%% file: decay_B3_cmp.tex
% GNUPLOT: LaTeX picture with Postscript
\begingroup
  \makeatletter
  \providecommand\color[2][]{%
    \GenericError{(gnuplot) \space\space\space\@spaces}{%
      Package color not loaded in conjunction with
      terminal option `colourtext'%
    }{See the gnuplot documentation for explanation.%
    }{Either use 'blacktext' in gnuplot or load the package
      color.sty in LaTeX.}%
    \renewcommand\color[2][]{}%
  }%
  \providecommand\includegraphics[2][]{%
    \GenericError{(gnuplot) \space\space\space\@spaces}{%
      Package graphicx or graphics not loaded%
    }{See the gnuplot documentation for explanation.%
    }{The gnuplot epslatex terminal needs graphicx.sty or graphics.sty.}%
    \renewcommand\includegraphics[2][]{}%
  }%
  \providecommand\rotatebox[2]{#2}%
  \@ifundefined{ifGPcolor}{%
    \newif\ifGPcolor
    \GPcolortrue
  }{}%
  \@ifundefined{ifGPblacktext}{%
    \newif\ifGPblacktext
    \GPblacktexttrue
  }{}%
  % define a \g@addto@macro without @ in the name:
  \let\gplgaddtomacro\g@addto@macro
  % define empty templates for all commands taking text:
  \gdef\gplbacktext{}%
  \gdef\gplfronttext{}%
  \makeatother
  \ifGPblacktext
    % no textcolor at all
    \def\colorrgb#1{}%
    \def\colorgray#1{}%
  \else
    % gray or color?
    \ifGPcolor
      \def\colorrgb#1{\color[rgb]{#1}}%
      \def\colorgray#1{\color[gray]{#1}}%
      \expandafter\def\csname LTw\endcsname{\color{white}}%
      \expandafter\def\csname LTb\endcsname{\color{black}}%
      \expandafter\def\csname LTa\endcsname{\color{black}}%
      \expandafter\def\csname LT0\endcsname{\color[rgb]{1,0,0}}%
      \expandafter\def\csname LT1\endcsname{\color[rgb]{0,1,0}}%
      \expandafter\def\csname LT2\endcsname{\color[rgb]{0,0,1}}%
      \expandafter\def\csname LT3\endcsname{\color[rgb]{1,0,1}}%
      \expandafter\def\csname LT4\endcsname{\color[rgb]{0,1,1}}%
      \expandafter\def\csname LT5\endcsname{\color[rgb]{1,1,0}}%
      \expandafter\def\csname LT6\endcsname{\color[rgb]{0,0,0}}%
      \expandafter\def\csname LT7\endcsname{\color[rgb]{1,0.3,0}}%
      \expandafter\def\csname LT8\endcsname{\color[rgb]{0.5,0.5,0.5}}%
    \else
      % gray
      \def\colorrgb#1{\color{black}}%
      \def\colorgray#1{\color[gray]{#1}}%
      \expandafter\def\csname LTw\endcsname{\color{white}}%
      \expandafter\def\csname LTb\endcsname{\color{black}}%
      \expandafter\def\csname LTa\endcsname{\color{black}}%
      \expandafter\def\csname LT0\endcsname{\color{black}}%
      \expandafter\def\csname LT1\endcsname{\color{black}}%
      \expandafter\def\csname LT2\endcsname{\color{black}}%
      \expandafter\def\csname LT3\endcsname{\color{black}}%
      \expandafter\def\csname LT4\endcsname{\color{black}}%
      \expandafter\def\csname LT5\endcsname{\color{black}}%
      \expandafter\def\csname LT6\endcsname{\color{black}}%
      \expandafter\def\csname LT7\endcsname{\color{black}}%
      \expandafter\def\csname LT8\endcsname{\color{black}}%
    \fi
  \fi
    \setlength{\unitlength}{0.0500bp}%
    \ifx\gptboxheight\undefined%
      \newlength{\gptboxheight}%
      \newlength{\gptboxwidth}%
      \newsavebox{\gptboxtext}%
    \fi%
    \setlength{\fboxrule}{0.5pt}%
    \setlength{\fboxsep}{1pt}%
\begin{picture}(4534.00,4534.00)%
    \gplgaddtomacro\gplbacktext{%
      \csname LTb\endcsname%
      \put(748,990){\makebox(0,0)[r]{\strut{}$0.01$}}%
      \csname LTb\endcsname%
      \put(748,1951){\makebox(0,0)[r]{\strut{}$0.1$}}%
      \csname LTb\endcsname%
      \put(748,2912){\makebox(0,0)[r]{\strut{}$1$}}%
      \csname LTb\endcsname%
      \put(748,3873){\makebox(0,0)[r]{\strut{}$10$}}%
      \csname LTb\endcsname%
      \put(1084,353){\rotatebox{70}{\makebox(0,0)[l]{\strut{}$2000$}}}%
      \csname LTb\endcsname%
      \put(1468,353){\rotatebox{70}{\makebox(0,0)[l]{\strut{}$4000$}}}%
      \csname LTb\endcsname%
      \put(1852,353){\rotatebox{70}{\makebox(0,0)[l]{\strut{}$6000$}}}%
      \csname LTb\endcsname%
      \put(2236,353){\rotatebox{70}{\makebox(0,0)[l]{\strut{}$8000$}}}%
      \csname LTb\endcsname%
      \put(2620,353){\rotatebox{70}{\makebox(0,0)[l]{\strut{}$10000$}}}%
      \csname LTb\endcsname%
      \put(3005,353){\rotatebox{70}{\makebox(0,0)[l]{\strut{}$12000$}}}%
      \csname LTb\endcsname%
      \put(3389,353){\rotatebox{70}{\makebox(0,0)[l]{\strut{}$14000$}}}%
      \csname LTb\endcsname%
      \put(3773,353){\rotatebox{70}{\makebox(0,0)[l]{\strut{}$16000$}}}%
    }%
    \gplgaddtomacro\gplfronttext{%
      \csname LTb\endcsname%
      \put(176,2431){\rotatebox{-270}{\makebox(0,0){\strut{}$\tau$}}}%
      \put(2508,154){\makebox(0,0){\strut{}$N$}}%
      \put(2508,4203){\makebox(0,0){\strut{}B3}}%
      \csname LTb\endcsname%
      \put(3150,3700){\makebox(0,0)[r]{\strut{}operator expansion}}%
      \csname LTb\endcsname%
      \put(3150,3480){\makebox(0,0)[r]{\strut{}real time}}%
    }%
    \gplbacktext
    \put(0,0){\includegraphics{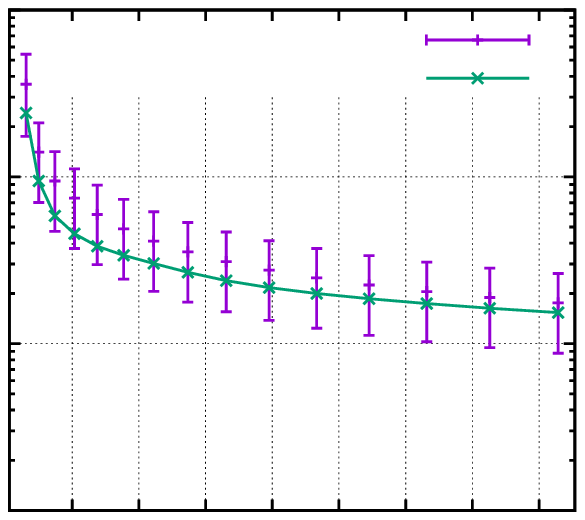}}%
    \gplfronttext
  \end{picture}%
\endgroup